\documentclass[12pt]{article}
\usepackage{amssymb,amsmath}
\usepackage[dvipdfmx]{graphicx}
\usepackage{graphicx}
\usepackage{subfigure}
\usepackage{ytableau}
\usepackage{comment}
\usepackage[vcentermath]{youngtab}
\usepackage{hyperref}
\usepackage{tikz}
\DeclareMathOperator*{\Tr}{{\rm Tr}}

\renewcommand{\thefootnote}{\fnsymbol{footnote}}

\numberwithin{equation}{section}

\begin{document}

%%%%%%%%%%%%%%%%%%%%%%%%%%%%%%%%%%%%%%%%%%%%
\thispagestyle{empty}
\begin{flushright}
RUP-24-9
\end{flushright}
%\vskip1cm
\vskip1cm
\begin{center}
{\bf {\LARGE Giant graviton expansions and ETW brane}}

\vskip1.5cm

Yasuyuki Hatsuda$^a$\footnote[0]{yhatsuda@rikkyo.ac.jp}, 

\bigskip
$^a$
{\it Department of Physics, Rikkyo University, \\
Toshima, Tokyo 171-8501, Japan
}

\bigskip
\bigskip

Hai Lin$^b$\footnote[0]{hailinhl@seu.edu.cn}

\bigskip
\bigskip

and 

\bigskip
\bigskip

Tadashi Okazaki$^b$\footnote[0]{tokazaki@seu.edu.cn}

\bigskip
$^b$
{\it Shing-Tung Yau Center of Southeast University,\\
Yifu Architecture Building, No.2 Sipailou, Xuanwu district, \\
Nanjing, Jiangsu, 210096, China
}

\end{center}
\renewcommand{\thefootnote}{\arabic{footnote}}

%%%%%%%%%%%%%%%%%%%%%%%%%%%%%%%%%%%%%%%%%%%%
\vskip1.25cm
\begin{abstract}
We study the giant gravitons in the $AdS_4$ bagpipe geometries involving end-of-the-world (ETW) brane 
constructed by a single $5$-brane and either two stacks or one stack of D3-branes in Type IIB string theory. 
From the exact formulae and giant graviton expansions of the half-indices for the half-BPS boundary conditions and interfaces in $\mathcal{N}=4$ super Yang-Mills theory, 
we obtain the BPS spectra of the fluctuation modes 
of the $AdS_4$ bagpipe geometries including the ETW brane region. 
\end{abstract}
%%%%%%%%%%%%%%%%%%%%%%%%%%%%%%%%%%%%%%%%%%

\newpage
\setcounter{tocdepth}{3}
\tableofcontents
%%%%%%%%%%%%%%%%%%%%%%%%%%

%%%%%%%%%%%%%%%%%%%%%%%%%%%%%%%%%%%%%%%%%%%%%%%
%%%%%%%%%%%%%%%%%%%%%%%%%%%%%%%%%%%%%%%%%%%%%%%
\section{Introduction and conclusion}
%%%%%%%%%%%%%%%%%%%%%%%%%%%%%%%%%%%%%%%%%%%%%%%
%%%%%%%%%%%%%%%%%%%%%%%%%%%%%%%%%%%%%%%%%%%%%%%
There are important motivations to study physical systems in the presence of boundaries. 
Boundary conditions in quantum field theories (QFTs) 
are necessary data to describe physical systems and phenomena across the dimension of spacetime. 
They are not only the conditions imposed on the bulk fields, but also admit an introduction of boundary degrees of freedom. 
Thus there are infinitely many boundary conditions as a generalized lower-dimensional system 
and they can shed light on fundamental properties of QFTs, such as cutting and gluing manipulations. 

The half-BPS boundary conditions in $\mathcal{N}=4$ $U(N)$ super Yang-Mills (SYM) theory are classified by Gaiotto and Witten \cite{Gaiotto:2008sa,Gaiotto:2008sd}. 
They can be realized in Type IIB string theory as a stack of $N$ D3-branes ending on and intersecting with $5$-branes \cite{Hanany:1996ie,Gaiotto:2008ak}. 
While $\mathcal{N}=4$ $U(N)$ SYM theory has a holographic dual geometry $AdS_5$ $\times$ $S^5$ in Type IIB theory, 
the holographic duals of the half-BPS boundary conditions in $\mathcal{N}=4$ SYM theory 
can provide a natural setup of double holography \cite{Karch:2000ct,Karch:2000gx,Takayanagi:2011zk,Karch:2022rvr} in string theory. 
D'Hoker, Estes and Gutperle constructed the supergravity solutions \cite{DHoker:2007zhm,DHoker:2007hhe} 
with the form of warped geometries corresponding to these interface structures. 
They realize the Karch-Randall models \cite{Karch:2000ct,Karch:2000gx} 
for which a part of the $AdS_5$ is cut off by an ETW brane with $AdS_4$ factor. 
The supergravity solutions have the shape of ``bagpipes'' 
where the bag describes a small perturbation to $AdS_4$ geometry while the pipes correspond to the highly-curved semi-infinite Janus throats \cite{Bachas:2018zmb}. 
They can provide remarkably useful examples for studying various new phenomena of ETW branes, 
\footnote{
%According to the cobordism conjecture \cite{McNamara:2019rup} 
%consistent theories of quantum gravity should include boundaries as configurations ending spacetime geometry, 
%which are identified with an effective end-of-the-world (ETW) brane. 
The ETW brane we focus on here is the additional boundary component of $AdS_5$ space in the bulk, that is different from the conformal boundary of $AdS_5$. 
}
including counting microstates, computing entanglement entropies, 
and describing phase transitions via the holography. 

In this paper we investigate the spectra of fluctuation modes on the $AdS_4$ bagpipe geometries with ETW brane 
which are holographically dual to the half-BPS boundary conditions in $\mathcal{N}=4$ $U(N)$ SYM theory 
and interfaces between $U(N)$ and $U(M)$ theories. 
The $AdS_4$ ETW brane can support giant gravitons 
as D3-branes wrapping a $3$-cycle formed by the fibration of the $S^2$ over a segment in the supergravity solutions.  
By analyzing the Lagrangian of the half-BPS giant graviton in the ETW brane, 
we find that there are upper bounds on the energies and angular momenta of the giant graviton 
in a similar manner as that wrapping $S^3$ in the $AdS_5\times S^5$ \cite{McGreevy:2000cw}. 
Subsequently, we obtain the giant graviton indices which capture the finite gauge rank corrections from the expansions of half-indices \cite{Dimofte:2011py,Gang:2012yr,Gaiotto:2019jvo,Okazaki:2019ony}, 
the supersymmetric indices which encode the spectrum of the boundary BPS local operators 
subject to the half-BPS boundary conditions. 
In particular, we show that 
the half-indices for the Neumann or Nahm pole boundary condition of $U(N)$ theory 
can be produced from the large $N$ normalized two-point functions of the Wilson lines transforming in the (anti)symmetric representations evaluated in \cite{Hatsuda:2023iwi,Hatsuda:2023imp}. 
Besides, we give an analytic proof of the recent observation of its inverse relation by Imamura and Inoue \cite{Imamura:2024pgp}.  
Therefore we analytically prove that the half-index of Neumann or Nahm pole boundary condition 
and the large $N$ normalized two-point functions of the Wilson lines transforming in the antisymmetric representation are dual with each other. 
Also we examine the half-indices for the 3d interfaces between $U(N)$ and $U(M)$ gauge theories constructed by a single $5$-brane. 
We obtain the conjectural closed-form formulae for arbitrary $N$ and $M$ involving multiple sums which generalize the results in \cite{Gaiotto:2019jvo}. For $N\ge M$ the formulae involve the multiple sums over $M$ non-negative indices. Remarkably, we find a new conjectural formula for the $U(N)$ Schur index in the similar form. 
We also obtain the giant graviton expansion of the interface half-indices which  
captures the finite gauge rank corrections of the indices associated to the upper bounds on the angular momenta of two types of giants as double sum expansions. 
It involves the double series which generalizes the giant graviton expansion of the half-indices for the Neumann or Nahm pole boundary condition. 
The sums are taken over a pair of non-negative integers $m$ and $k$  
where $m$ corresponds to the wrapping number for the giants in the asymptotic $AdS_5\times S^5$ region and $k$ the wrapping number for the giants in the ETW region. 
Furthermore, the double sum brane expansion of the two-point function of the antisymmetric Wilson lines in \cite{Imamura:2024pgp} can be viewed as the dual relation. 

The paper is organized as follows. 
In section \ref{sec_IIBdual} we begin by reviewing the Type IIB supergravity solutions involving the ETW brane \cite{DHoker:2007zhm,DHoker:2007hhe}
which are dual to the half-BPS boundary conditions and interfaces in $\mathcal{N}=4$ SYM theory. 
We then present semiclassical analysis of a giant graviton in the ETW brane as a wrapped D3-brane. 
We derive the upper bounds of its energy and angular momentum. 
In section \ref{sec_hindex} we study the half-indices of the half-BPS boundary conditions and interfaces in $\mathcal{N}=4$ SYM theory. 
The Kaluza-Klein (KK) modes on the $AdS_4$ bagpipe geometries involving the ETW brane as well as the asymptotic $AdS_5\times S^5$ regions  
are obtained from the large $N$ limit of the half-indices. 
We find that they admit giant graviton expansions which control their finite gauge rank corrections. 

%%%%%%%%%%%%%%%%%%%%%%%%%%%%%%%%%%%%%%%%%%%%%%%
%%%%%%%%%%%%%%%%%%%%%%%%%%%%%%%%%%%%%%%%%%%%%%%
\section{Type IIB gravity dual}
\label{sec_IIBdual}
%%%%%%%%%%%%%%%%%%%%%%%%%%%%%%%%%%%%%%%%%%%%%%%
%%%%%%%%%%%%%%%%%%%%%%%%%%%%%%%%%%%%%%%%%%%%%%%
Here we first review the Type IIB supergravity solutions 
constructed by D'Hoker, Estes and Gutperle \cite{DHoker:2007zhm,DHoker:2007hhe} 
(also see \cite{Aharony:2011yc,Assel:2011xz} for their holographic interpretations  
and \cite{Raamsdonk:2020tin,VanRaamsdonk:2021duo,Demulder:2022aij,Huertas:2023syg} for their applications) which 
are dual to the half-BPS boundary conditions and interfaces in $\mathcal{N}=4$ $U(N)$ SYM theory. 
The supergravity solutions are the warped geometries as fibrations over a Riemann surface. 
They are described by ETW branes on which the $AdS_5\times S^5$ terminate. 
We make a semiclassical analysis of the giant gravitons as wrapped D3-branes in the ETW branes. 
In particular, the half-BPS giants wrap three-cycles appearing as fibrations of two-spheres over a segment on the boundary of the Riemann surface. 

%%%%%%%%%%%%%%%%%%%%%%%%%%%%%%%%%%%%%%%%%%%%%%%
\subsection{Brane setup}
%%%%%%%%%%%%%%%%%%%%%%%%%%%%%%%%%%%%%%%%%%%%%%%
The half-BPS boundary conditions in $\mathcal{N}=4$ $U(N)$ SYM theory can be realized in Type IIB string theory as the following Hanany-Witten brane configuration \cite{Hanany:1996ie}: 
\begin{align}
\label{branesetup}
\begin{array}{c|cccccccccc} 
&0&1&2&3&4&5&6&7&8&9 \\ \hline
\textrm{D3}&\circ&\circ&\circ&&&&\circ&&& \\
\textrm{NS5}'&\circ&\circ&\circ&&&&&\circ&\circ&\circ \\
\textrm{D5}'&\circ&\circ&\circ&\circ&\circ&\circ&&&& \\
\end{array}. 
\end{align}
Here $\circ$ stands for the direction along which the corresponding branes extend. 
The configuration involves a stack of $N$ D3-branes whose low-energy description is $\mathcal{N}=4$ $U(N)$ SYM theory. 
The boundary of $\mathcal{N}=4$ SYM theory is supported along the $012$ directions which are shared by both D3-branes and $5$-branes. 
The configuration (\ref{branesetup}) preserve the $SO(2,3)$ $\times$ $SO(3)$ $\times$ $SO(3)$ global symmetry group, 
where the factor $SO(2,3)$ is the three-dimensional conformal group 
and the factor $SO(3)\times SO(3)$ corresponds to the R-symmetry group. 
We call the $SU(2)$ $\cong$ $SO(3)$ associated with the $345$ directions 
and that with the $789$ directions $SU(2)_C$ and $SU(2)_H$ respectively. 
\footnote{
We follow the convention in \cite{Gaiotto:2019jvo} 
where the brane configuration with a pair of the NS5-branes extending along the $012345$ directions 
and the D5-branes along the $012789$ directions leads to the 3d $\mathcal{N}=4$ ordinary supermultiplets, 
while a pair of NS5$'$-barnes and D5$'$-branes in (\ref{branesetup}) admits the twisted supermultiplets 
for which the isometry of the Coulomb branch is $SU(2)_H$ and that of the Higgs branch is $SU(2)_C$. 
}
It is the bosonic subgroup of the three-dimensional superconformal group $OSp(2,2|4)$ $\subset$ $PSU(2,2|4)$. 

%%%%%%%%%%%%%%%%%%%%%%%%%%%%%%%%%%%%%%%%%%%%%%%
\subsection{ETW brane geometry}
\label{sec_ETW}
%%%%%%%%%%%%%%%%%%%%%%%%%%%%%%%%%%%%%%%%%%%%%%%
In the near-horizon limit of the brane configuration (\ref{branesetup}), 
the holographically dual geometry as a solution in Type IIB supergravity has the $AdS_4$ factor associated with the $SO(2,3)$ conformal group 
and the $S^2\times S^2$ factor accommodated by the $SO(3)\times SO(3)$ R-symmetry group. 
The holographically dual geometry in Type IIB supergravity takes the form \cite{DHoker:2007zhm,DHoker:2007hhe} 
\begin{align}
AdS_4\times S^2_{(1)}\times S^2_{(2)} \times \Sigma, 
\end{align}
that is the $AdS_4$ $\times$ $S^2$ $\times$ $S^2$ fibration over the Riemann surface $\Sigma$ with the topology of a disk. 
Here we denote the two types of spheres by $S_{(1)}^{2}$ and $S_{(2)}^{2}$. 

The 10d metric of the bulk geometries takes the form
\begin{align}
\label{metric_ansatz}
ds^{2}=f_{4}^{2}ds_{\mathrm{AdS}%
_{4}}^{2}+f_{1}^{2}ds_{S_{(1)}^{2}}^{2}+f_{2}^{2}ds_{S_{(2)}^{2}}^{2}+4\rho
^{2}d\bar{w}dw,
\end{align}%
The $ds_{\mathrm{AdS_{4}}}^{2}$ are line elements of $AdS_{4}$, and 
\begin{align}
ds_{S^2_{(1)}}^2&=d\theta^2+\cos^2\theta d\varphi^2, \\
ds_{S^2_{(2)}}^2&=d\chi^2+\cos^2\chi d\psi^2. 
\end{align}
are line elements of the two types of $S^{2}$. 
The warp factors are given by the harmonic functions $h$ and $\hat{h}$ on $\Sigma $,
\begin{align}
\label{f1f2}
f_{1}^{8} &=16h^{8}\frac{N_{2}W^{2}}{N_{1}^{3}},& 
f_{2}^{8}&=16\hat{h}^{8}\frac{N_{1}W^{2}}{N_{2}^{3}},\\
\label{f4rho}
f_{4}^{8}&=16\frac{N_{1}N_{2}}{W^{2}},&  
\rho ^{8}&=\frac{N_{1}N_{2}W^{2}}{h^{4}\hat{h}^{4}},
\end{align}
where
\begin{align}
N_{1}&=2h\hat{h}|\partial_w h|^{2}-h^{2}W,\\
N_{2}&=2h\hat{h}|\partial_w \hat{h}|^{2}-\hat{h}^{2}W, \\
W&=\partial_{w}\partial _{{\bar{w}}}(h\hat{h}).
\end{align}%
Any solution can map to the one with vanishing axion 
according to the $SL(2,\mathbb{R})$ symmetry of Type IIB supergravity. 
The dilaton is given by 
\begin{align}
e^{-2\phi }=\sqrt{\frac{N_{1}}{N_{2}}}.
\end{align}
There exist non-trivial backgrounds for the NSNS and RR $3$-forms and RR $5$-form. 
The RR and NSNS $3$-forms can be written in terms of the potentials 
\begin{align}
C^{(2)}&=b_2\omega^{89},\\
B^{(2)}&=b_1\omega^{45}, 
\end{align}
where 
\begin{align}
b_2&=2i h\frac{h\hat{h}(\partial h\bar{\partial}\hat{h}-\bar{\partial}h\partial \hat{h})}
{N_1}-2(A_1+\bar{A}_1), \\
b_1&=2i\hat{h}\frac{h\hat{h}(\partial h\bar{\partial}\hat{h}-\bar{\partial}h\partial\hat{h})}
{N_2}+2i(A_2-\bar{A}_2),
\end{align}
where $A_1$ and $A_2$ are defined in terms of the harmonic functions $h$ and $\hat{h}$ by
\begin{align}
h&=-i(A_1-\bar{A}_1), \\
\hat{h}&=A_2+\bar{A}_2. 
\end{align}
$\omega^{45}$ and $\omega^{89}$ are the volume forms of the two-spheres $S^{2}_{(1)}$ and $S^2_{(2)}$. 

The RR 5-form is 
\begin{align}
F^{(5)}=(1+\ast )(-4\,f_{4}^{\,4}\,\omega ^{\,0126}\wedge {F}), 
\end{align}
where $\omega^{0126}$ is the volume form of the $AdS_4$, 
$F$ is a $1$-form on $\Sigma$ 
and $\ast$ denotes the Poincare dual with respect to the 10d metric. 
Note that $F$ can be written as
\begin{align}
\ f_{4}^{\,4}{F}=dj
\end{align}
where 
\begin{align}
j=3{C}+3\bar{{C}}-3{D}
+i\frac{h\hat{h}}{W}\,(\partial h\bar{\partial}\hat{h}-\bar{\partial}h\partial \hat{h}). 
\end{align}
Here $C$ and $D$ are 
\begin{align}
\partial C&=A_1\partial A_2-A_2\partial A_1, \\
D&=\bar{A}_1 A_2+A_1 \bar{A}_2. 
\end{align}

Concrete supergravity solutions are determined by a choice of $\Sigma$ and a pair of two harmonic functions $h$ and $\hat{h}$ on $\Sigma$. 
The functions $h$ and $\hat{h}$ can be defined on an auxiliary hyperelliptic Riemann surface of genus $g$ 
as a compactification of the hyperelliptic curve
\begin{align}
s(u)^2&=(u-e_1)\prod_{i=1}^{g}(u-e_{2i})(u-e_{2i+1}). 
\end{align}
It can be taken as the lower half-plane 
with the branch points $e_i$ $\in$ $\mathbb{R}$. 
The $SL(2,\mathbb{R})$ symmetry of the lower half-plane is used to set 
the branch point $e_{2g+2}$ to $+\infty$. 

The functions $h$ and $\hat{h}$ are analytic in the interior of $\Sigma$, however, they can have singularities on its boundary 
in such a way that the differentials $\partial h$ and $\partial \hat{h}$ have poles on boundary of $\Sigma$. 
The holomorphic differentials take the forms
\begin{align}
\partial h&=-i\frac{P(u)Q_1(u)}{s(u)^3} du, \\
\partial \hat{h}&=-\frac{P(u)Q_2(u)}{s(u)^3} du, 
\end{align}
where 
\begin{align}
P(u)&=\prod_{i=1}^{g}(u-u_i)(u-\overline{u}_i), \\
Q_{1}(u)&=\prod_{j=1}^{g+1}(u-\alpha_j), \\
Q_2(u)&=\prod_{j=1}^{g+1}(u-\beta_j). 
\end{align}
$P(u)$ is a real polynomial in $u$ of degrees $2g$ with complex zeros 
and $Q_{i}(u)$ are real polynomials in $u$ of degrees $g+1$ with real zeros.  

According to the regularity conditions, 
the zeros of polynomials $P(u)$, $Q_i(u)$ and branch points should be aligned on a real axis as follows: 
\begin{align}
\alpha_{g+1}\le e_{2g+2}\le \beta_{g+1}\le e_{2g}\le \cdots 
\le \alpha_{j}\le e_{2j-1}\le \beta_j\le \cdots
\le e_2\le \alpha_1\le e_1\le\beta_1. 
\end{align}

The functions $h$ and $\hat{h}$ satisfy alternating Neumann and vanishing Dirichlet boundary conditions, 
i.e. $h$ obeys the Neumann boundary condition whenever $\hat{h}$ satisfies the Dirichlet boundary condition vice versa. 
The boundary conditions change when one crosses the branch points $e_i$. 
Since the values of one of the $h$ and $\hat{h}$ vanish on each interval, we have
\begin{align}
\int_{e_{2i}}^{e_{2i-1}}(\partial_u h du+\partial_{\bar{u}}h d\bar{u})&=0, \\
\int_{e_{2i+1}}^{e_{2i}}(\partial_u \hat{h} du+\partial_{\bar{u}}\hat{h} d\bar{u})&=0, 
\end{align}
from which the values of $\{u_i\}$ can be determined in terms of $\{\alpha_i\}$, $\{\beta_i\}$ and $\{e_i\}$. 

The boundary conditions of $h$ and $\hat{h}$ encode the cycles in the full geometry. 
The $S^2_{(2)}$ (resp. $S^2_{(1)}$) metric factors shrink to zero size on the segments $(e_{2i+1},e_{2i})$ 
(resp. $(e_{2i},e_{2i-1})$) on the boundary of the Riemann surface 
whereas the other factors $S^2_{(1)}$ (resp. $S^2_{(2)}$) are finite. 
Consequently, non-trivial $3$-cycles $C_{(2)i}$ (resp. $C_{(1)i}$) arise as fibrations of the $2$-sphere over the intervals
\begin{align}
\label{3cycle_c1}
C_{(1)i}&=\left\{ [e_{2i},e_{2i-1}]\times_f S_{(1)}^2 \right\}, \\
\label{3cycle_c2}
C_{(2)i}&=\left\{ [e_{2i+1},e_{2i}]\times_f S_{(2)}^2 \right\}, 
\end{align}
where $C_{(j)}$ is topologically a $3$-sphere. 
Also we have non-trivial $5$-cycles as fibrations of the two $2$-spheres over different types of the intervals separated by one branch point. 
The compact part of the geometry will have the topology of $S^5$. 
Therefore the boundary of the Riemann surface is not an actual boundary of the geometry. 

%%%%%%%%%%%%%%%%%%%%%%%%%%%%%%%%%%%%%%%%%%%%%%%
\subsubsection{Asymptotic $AdS_5\times S^5$ regions}
%%%%%%%%%%%%%%%%%%%%%%%%%%%%%%%%%%%%%%%%%%%%%%%
Let us choose coordinates $v=u-e_i$ for $i=1,\cdots, 2g+1$ and $v=1/u$ for $i=2g+2$. 
Near the branch point $e_i$ he harmonic function $h$ and $\hat{h}$ have the form 
\cite{DHoker:2007hhe}
\begin{align}
h&=2i \left( \gamma^i \frac{1}{\sqrt{v}}-\delta^i \sqrt{v} \right)+\mathcal{O}(v^{3/2})+c.c., \\
\hat{h}&=2\left( \hat{\gamma}^i \frac{1}{\sqrt{v}}-\hat{\delta}^i \sqrt{v} \right)+\mathcal{O}(v^{3/2})+c.c. 
\end{align}
where 
\begin{align}
\gamma^i&=\frac{P(e_i)Q_1(e_i)}{\prod_{j\neq i}(e_i-e_j)^{3/2}}, \\
\hat{\gamma}^i&=\frac{P(e_i)Q_2(e_i)}{\prod_{j\neq i}(e_i-e_j)^{3/2}}, 
\end{align}
\begin{align}
\delta^i&=\gamma^i \left[
\sum_{k=1}^{g}
\left( \frac{1}{e_i-u_k}+\frac{1}{e_i-\bar{u}_k} \right)
+\sum_{k=1}^{g+1}\frac{1}{e_i-\alpha_k}
-\frac32\sum_{k\neq i}^{2g+1}\frac{1}{e_i-e_k}
\right], \\
\hat{\delta}^i&=\hat{\gamma}^i \left[
\sum_{k=1}^{g}
\left( \frac{1}{e_i-u_k}+\frac{1}{e_i-\bar{u}_k} \right)
+\sum_{k=1}^{g+1}\frac{1}{e_i-\beta_k}
-\frac32\sum_{k\neq i}^{2g+1}\frac{1}{e_i-e_k}
\right]
\end{align}
for $i=1,\cdots 2g+1$ 
and 
\begin{align}
\gamma^{2g+2}&=i, \\
\hat{\gamma}^{2g+2}&=i, \\
\delta^{2g+2}&=i\sum_{k=1}^{g}(u_k+\bar{u}_k)
+\sum_{k=1}^{g+1}\alpha_k-\frac32 \sum_{k=1}^{2g+1}e_k, \\
\hat{\delta}^{2g+2}&=i\sum_{k=1}^{g}(u_k+\bar{u}_k)
+\sum_{k=1}^{g+1}\beta_k-\frac32 \sum_{k=1}^{2g+1}e_k
\end{align}
for $i=2g+2$. 

Furthermore, introducing new coordiantes $v=e^{-2x-2iy}$ 
with $-\infty \le x\le \infty$ and $0\le y\le \pi/2$, 
the asymptotic region near the branch point $e_i$ at $v=0$ corresponds to $x\rightarrow \infty$. 

Then the dilaton is given by
\begin{align}
\label{asymp_ads5s5}
e^{-2\phi}&\sim \frac{\gamma^i}{\hat{\gamma}^i}+\mathcal{O}(e^{-4x})
\end{align}
and the metric becomes 
\begin{align}
ds^2&=4\sqrt{2\Delta^i}
(e^{2x}ds_{AdS_4}^2+(\cos y)^2 ds_{S_{(1)}^2}^2+(\sin y)^2 ds_{S_{(2)}^2}^2+dx^2+dy^2)
+\mathcal{O}(e^{-2x})
\end{align}
where 
\begin{align}
\Delta^i&=\gamma^i \hat{\delta}^i-\hat{\gamma}^i \delta^i. 
\end{align}
In the limit $x\rightarrow \infty$, the vicinity of the branch point $e_i$, 
(\ref{asymp_ads5s5}) becomes the metric of the $AdS_5\times S^5$ 
with the radius squared $4\sqrt{2 \Delta^i}$. 
Therefore near each branch point $e_i$ the Riemann surface develops a semi-infinite spike 
corresponding to the asymptotic $AdS_5$ $\times$ $S^5$ 
so that the genus $g$ solution contains $2g+2$ asymptotic $AdS_5\times S^5$ regions. 
They stick out of the $AdS_4\times X_6$ geometry where $X_6$ is a compact six-dimensional manifold. 
The resulting geometries are referred to as ``bagpipes'' \cite{Bachas:2018zmb}. 
The $5$-form flux over the $5$-sphere near the branch point $e_i$ is evaluated as
\begin{align}
\lim_{x\rightarrow \infty}\int_{S^5}
F^{(5)}&=2 (4\pi)^3 \Delta^i
\end{align}
and 
\begin{align}
N_{\textrm{D3}_i}&=8(4\pi)^3 |\Delta^i|\in \mathbb{Z}
\end{align}
is the number of D3-branes associated with the asymptotic $i$-th $AdS_5\times S^5$ region.   

However, the genus $g>0$ solutions that contain interior zeros $\{u_i\}$ are singular at $u=u_i$ 
due to the conical singularity of the Riemann surface $\Sigma$. 
Accordingly, generic non-singular solutions are the genus $g=0$ solutions, in the hyperelliptic formulation. 
Nevertheless, the higher genus $g>0$ solutions admit degeneration limits  
which can describe $g$ stacks of the $5$-branes. 
When a subset $(e,\alpha,e)$ collapses, 
the geometry involves a stack of the NS5$'$-branes at the degeneration point. 
On the other hand, when a subset $(e,\beta,e)$ collapses, 
the geometry has a stack of the D5$'$-branes at the degeneration point.  
After the collapses of $g$ branch points, 
the geometry has $g$ stacks of the $5$-branes 
and two asymptotic $AdS_5\times S^5$ regions 
which support the $5$-form fluxes corresponding to 
the two half-spaces of 4d $\mathcal{N}=4$ SYM theories. 
In the following, we focus on the genus $g=1$ solutions 
which have a single stack of the $5$-branes in the degeneration limit. 

%%%%%%%%%%%%%%%%%%%%%%%%%%%%%%%%%%%%%%%%%%%%%%%
\subsubsection{The genus $g=1$ solutions}
%%%%%%%%%%%%%%%%%%%%%%%%%%%%%%%%%%%%%%%%%%%%%%%
For genus $g=1$ solution
there are four branch points at $u=e_{1,2,3}$ and $e_4=\infty$ 
as well as the zeros $\alpha_{1,2}$ of $\partial h$ and those $\beta_{1,2}$ of $\partial \hat{h}$
\begin{align}
\label{collapse_1}
\alpha_2 < e_3 < \beta_2 < e_2 < \alpha_1 < e_1 < \beta_1. 
\end{align}
There are two degeneration limits corresponding to the single stack of D5$'$-branes and that of NS5$'$-branes. 
Let us consider the solutions involving the stack of D5$'$-branes 
for which the subset $(e_3 \beta_2 e_2)$ collapses. 
It can be achieved when we set $e_3=e_2=\beta_2=-l^2$. 

Take $e_1=0$ and relabeling $\beta_1=\beta$, we have
\begin{align}
\label{collapse_2}
\alpha_2 < -l^2 < \alpha_1 < 0 < \beta. 
\end{align}

From the regularity and negativity of the function $W$ 
a complex zero $u_1$ is determined as $u_1=-l^2$. 
We obtain the differentials
\begin{align}
\label{diff_g1}
\partial h&=-i \frac{(u-\alpha_1)(u-\alpha_2)}{(u+l^2)u^{\frac32}}du,\\
\label{diff_g2}
\partial \hat{h}&=-\frac{u-\beta}{u^{\frac32}}du. 
\end{align}
The $(\frac32,\frac32)$ singularities at $u=0, \infty$ correspond to the two asymptotic $AdS_5\times S^5$ regions. 
The $(1,0)$ singularity at $u=-l^2$ corresponds to the stack of D5$'$-branes. 
The solution near the branch points at $e_1=0$ and $e_4=\infty$ is encoded by 
\begin{align}
\gamma^1&=\frac{\alpha_1\alpha_2}{l^2},\qquad 
\hat{\gamma}^1=-\beta,\qquad 
\gamma^4=i, \qquad 
\hat{\gamma}^4=-i, \\
\delta^1&=-\frac{l^2(\alpha_1+\alpha_2)+\alpha_1\alpha_2}{l^4}, \qquad 
\hat{\delta}^1=1,\qquad
\delta^4=i(\alpha_1+\alpha_2+l^2), \qquad 
\hat{\delta}^4=i\beta. 
\end{align}
The D3-brane charges in the two asymptotic $AdS_5\times S^5$ regions are given by
\begin{align}
M&:=N_{\textrm{D3}_1}=8(4\pi)^3 |\Delta^1|\in \mathbb{Z}, \\
N&:=N_{\textrm{D3}_4}=8(4\pi)^3 |\Delta^4|\in \mathbb{Z}, 
\end{align}
with 
\begin{align}
\Delta^1&=
\frac{-\beta l^2(\alpha_1+\alpha_2)-\beta\alpha_1\alpha_2 +\alpha_1\alpha_2 l^2}{l^4}, \\
\Delta^4&=
-\alpha_1-\alpha_2-\beta-l^2. 
\end{align}

We choose a new coordinate $w^2=u$, 
which parametrizes the second quadrant in the complex plane
\begin{align}
\Sigma&=\{w| \Re(w)\le 0, \Im(w)\ge0\}. 
\end{align}
The differentials (\ref{diff_g1}) and (\ref{diff_g2}) can be integrated 
so that the harmonic functions are given by
\begin{align}
h&=-2i (w-\bar{w})\left(1-\frac{\alpha_1\alpha_2}{l^2|w|^2}\right)
-\frac{(\alpha_1+l^2)(l^2+\alpha_2)}{l^3}\ln \frac{|w+il|^2}{|w-il|^2}, \\
\hat{h}&=-2(w+\bar{w})\left(1+\frac{\beta}{|w|^2}\right). 
\end{align}
Introducing new coordinates $(\xi,\zeta)$ defined by $w=il+\xi e^{i\zeta}$, 
one can see the expansion around the collapsed branch points at $w=il$. 
Then the metric of the Riemann surface can be written as
\begin{align}
d\bar{w}dw&=d\xi^2+\xi^2 d\zeta^2. 
\end{align}
Dropping the subleading terms in $\xi$ and $\log \xi$, 
the harmonic functions can be expanded as
\begin{align}
h&=2\tilde{b}+2\tilde{d}\ln (\xi^2), \\
\hat{h}&=-4 \tilde{c} \xi \cos\zeta, 
\end{align}
where 
\begin{align}
\tilde{c}&=1+\frac{\beta}{l^2}, \\
\tilde{b}&=2l-\frac{2\alpha_1\alpha_2}{l^3}
-\frac{(\alpha_1+l^2)(l^2+\alpha_2)}{l^3}\ln (4l^2), \\
2\tilde{d}&=\frac{(\alpha_1+l^2)(l^2+\alpha_2)}{l^3}. 
\end{align}
In this limit the solution becomes 
\begin{align}
e^{-4\phi}&\sim \frac{\tilde{d}^2}{\tilde{c}^2}\frac{\log\xi}{\xi^2}, \\
\rho^2&\sim \frac{2\sqrt{\tilde{c}\tilde{d}}}{\xi^{\frac32}|\ln \xi|^{\frac14}}, \\
f_1^2&\sim 8\sqrt{\tilde{c}\tilde{d}}\xi^{\frac12}|\ln \xi|^{\frac34}, \\
f_2^2&\sim 8\sqrt{\tilde{c}\tilde{d}}\frac{\xi^{\frac12} (\cos\zeta)^2}{|\ln \xi|^{\frac14}}, \\
f_4^2&\sim 8\sqrt{\tilde{c}\tilde{d}}\xi^{\frac12}|\ln \xi|^{\frac34}. 
\end{align}
Therefore near the location of the D5$'$-brane we obtain the metric of the form
\begin{align}
\label{ETW_D5}
ds^2&=f_4(\xi)^2
\left( ds_{AdS_4}^2+ds_{S_{(1)}^2}^2 \right)
+4\xi^2\rho(\xi)^2
\left(
d\zeta^2+\cos^2\zeta ds_{S_{(2)}^2}^2
+\frac{1}{\xi^2}d\xi^2
\right). 
\end{align}
For fixed $\xi$ the metric includes 
the $AdS_4\times S_{(1)}^2$ factor that corresponds to the world-volume of a stack of D5$'$-branes. 
Also it has the $3$-cycle $C_{(2)}$, defined by (\ref{3cycle_c2}), which supports the RR $3$-form flux given by
\begin{align}
\int_{C_{(2)}} F^{(3)}
&=32\pi^2 \tilde{d}=N_{\textrm{D5}}\in \mathbb{Z}
\end{align}
whereas the NSNS $3$-form flux vanishes. 
The solution has a stack of $N_{\textrm{D5}}$ D5$'$-branes wrapping $AdS_4\times S_{(1)}^2$. 

The conserved $5$-form flux associated with the D5$'$-brane singularity at $w=il$ is evaluated as
\begin{align}
\label{D3ch_D5}
\int_{S_{(1)}^2\times C_{(2)}}
\left( F^{(5)}-B^{(2)} \wedge F^{(3)} \right)
=N_{\textrm{D3}}^{\textrm{D5}}. 
\end{align}
where
\begin{align}
\label{D3ch_D5a}
N_{\textrm{D3}}^{\textrm{D5}}&
=N_{\textrm{D3}_4}-N_{\textrm{D3}_1}
=N-M
\end{align}
is interpreted as the number of D3-branes ending on the stack of D5-branes. 

%%%%%%%%%%%%%%%%%%%%%%%%%%%%%%%%%%%%%%%%%%%%%%%
\subsection{Giant gravitons and ETW brane}
\label{sec_gg_ETW}
%%%%%%%%%%%%%%%%%%%%%%%%%%%%%%%%%%%%%%%%%%%%%%%
%Giant gravitons
While the gravitons in the near horizon geometry 
produced by branes are point-like at low angular momentum, 
those at large angular momentum behave as extended branes. 
They are called giant gravitons \cite{Myers:1999ps,McGreevy:2000cw,Grisaru:2000zn,Hashimoto:2000zp}. 

%%%%%%%%%%%%%%%%%%%%%%%%%%%%%%%%%%%%%%%%%%%%%%%
\subsubsection{Giant gravitons in $AdS_5\times S^5$}
%%%%%%%%%%%%%%%%%%%%%%%%%%%%%%%%%%%%%%%%%%%%%%%
%1/2BPS giant gravitons
For the $AdS_5\times S^5$ geometry obtained as near horizon limit of a stack of $N$ D3-branes, 
gravitons of energies at order $N$ blow up into puffed up D3-branes as giant gravitons. 
There exist two types of half-BPS giant gravitons. 
The first one is identified with spherical D3-branes wrapping time direction in $AdS_5$ and $S^3$ $\subset$ $S^5$. 
They behave as a particle in $AdS_5$. See also \cite{Corley:2001zk,Lin:2004nb,Berenstein:2004kk,Balasubramanian:2001nh} for discussions on related holographic aspects of giants. 
The other is D3-branes wrapping $S^3$ $\subset$ $AdS_5$ and an equator on $S^5$, the dual giant gravitons \cite{Grisaru:2000zn,Hashimoto:2000zp}. 
Here we focus only on the giant gravitons, while it may be an interesting future direction to explore the index calculation with dual giant gravitons. 

%1/8BPS giant gravitons
Besides, a large class of $1/8$-BPS giant gravitons in the $AdS_5\times S^5$ is constructed by Mikhailov \cite{Mikhailov:2000ya}. 
Such giant gravitons appear as D3-branes wrapping 
holomorphic cycles defined by the intersection of the zero set of holomorphic function in $\mathbb{C}^3$ with $S^5$. 

%%%%%%%%%%%%%%%%%%%%%%%%%%%%%%%%%%%%%%%%%%%%%%%
\subsubsection{Giant gravitons in ETW brane}
%%%%%%%%%%%%%%%%%%%%%%%%%%%%%%%%%%%%%%%%%%%%%%%
%giant gravitons in ETW brane
Now we consider a similar brane configuration of giant gravitons in the $AdS_4$ ETW brane region. 
Since the $3$-cycle $C_{(i)}$ which is the fibration of $S_{(i)}^{2}$, $i=1,2$ over the segment is topologically a three-sphere, 
we can introduce extra D3-branes wrapping this cycle as the half-BPS giant gravitons. 
They have a spherical symmetry as rotations of the world-volume. 

Consider a giant graviton D3-brane wrapping the $C_{(2)}$ formed by $S_{(2)}^{2}$ fibration over a segment on the boundary of the Riemann surface 
sitting at $w=il + \xi e^{i\zeta}$ inside the ETW brane near the location of the D5$'$-brane at $w=il$ where $il$ is defined in (\ref{collapse_2}). 

When we introduce the coordinates $(\sigma_0,\sigma_i)$ with $i=1,2,3$ on the giant graviton, 
in the static gauge the time coordinate $\sigma _{0}$ on the giant graviton is identified with the time in $AdS_{4}$ 
and the spatial coordinates $\sigma _{i},i=1,2,3$ of the giant are identified with $x$ and the coordinates $(\chi,\psi)$ on the two-sphere $S_{(2)}^{2}$
\begin{align}
\sigma_0&=t, &
\sigma_1&=\zeta, & 
\sigma_2&=\chi, &
\sigma_3&=\psi. 
\end{align}
The giant graviton has dynamical motions in $AdS_{4}\times S_{(1)}^{2}$. 
As we will see, it behaves, analogously, as a particle with the quantized angular momentum 
moving along the angle $\varphi$ in $S_{(1)}^{2}$. 
This angular momentum corresponds to the $U(1)$ charge, 
that is the Cartan of the $SU(2)_{C}$ in the Coulomb limit of the index. 

%ansatz for GG action
We make the following ansatz for the embedding of the D3-brane giant: 
\begin{align}
\label{ansatz_1/2}
\varphi&=\varphi(t),&
\theta&=\theta(t),&  
\xi&=\textrm{const}.   
\end{align}
%ETW geometry (pullback metric)
The pull-back of the geometry (\ref{ETW_D5}) near the location of the D5$'$-brane is 
\begin{align}
\label{pullback_g}
g_{mn}&=
\left(
\begin{matrix}
-f_4^2+f_4^2\cos^2\theta \dot{\varphi}^2+f_4^2\dot{\theta}^2
&0\\
0&4\xi^2 \rho^2 g_{ij}\\
\end{matrix}
\right), 
\end{align}
where $g_{ij}$ stands for the metric of the $3$-cycle $C_{(2)}$ 
\begin{align}
ds^2&=d\zeta^2+\cos^2 \zeta ds_{S_{(2)}^2}^2. 
\end{align}

%ETW geometry (4-form)
Since the $5$-cycles $S_{(1)}^2\times C_{(2)}$ has total conserved flux quanta (\ref{D3ch_D5})
in the region near $w=il$, 
we have the $4$-form potential 
\begin{align}
\label{pullback_A}
A^{(4)}
&=N_{\textrm{D3}}^{\textrm{D5}} \sin \theta d\varphi \cos ^{2}\zeta d\zeta d\Omega _{S_{(2)}^{2}}. 
\end{align}
One can locally write 
\begin{align}
\cos^2\zeta d\zeta d\Omega_{S^2_{(2)}}&=d\Omega_{3}. 
\end{align}

%action of D3-brane
The bosonic action of the D3-brane is
\begin{align}
\label{D3_action}
S&=-\mu_3 \int d^4\sigma \sqrt{-g}+\mu_3\int P[A^{(4)}], 
\end{align}
where $\mu_3$ is the D3-brane tension, $g_{mn}$ is the pull-back of the spacetime metric and $P[A^{(4)}]$ is the pull-back of the 4-form potential. 

%Lagrangian of GG
We denote the Lagrangian density of the brane by $\mathcal{L}$, 
where the D3-brane action is $S=\mu _{3}\int \mathcal{L}dtd^{3}\Omega$. For the half-BPS giant graviton, substituting the solution (\ref{ansatz_1/2}), the pull-backs (\ref{pullback_g}) and (\ref{pullback_A}) into the action (\ref{D3_action}) 
and integrating over the spatial coordinates on the world-volume, we can reduce the action to the Lagrangian $L$ in the $(0+1)$ dimensional quantum mechanics, 
$S=\mu _{3}V_{\Omega_{3}}\int Ldt$ where $V_{\Omega_3}$ is the volume of the $3$-cycle. 
We find 
\begin{align}
\label{Lag_ETWGG1/2}
L&=
-64\tilde{c}\tilde{d} \xi 
\sqrt{1-\cos ^{2}\theta \dot{\varphi}^{2}-\dot{\theta}^2}
+N_{\textrm{D3}}^{\textrm{D5}} \sin\theta \dot{\varphi}, 
\end{align}
where the dot is the time derivative. 
The expression (\ref{Lag_ETWGG1/2}) is the Lagrangian of the half-BPS giant graviton in the $AdS_4$ ETW brane near the single stack of D5$'$-branes at $w=il$. 

%angular momentum
The angular momentum is
\begin{align}
\label{Pvarphi}
P_{\varphi}&=\frac{64\tilde{c}\tilde{d}\cos^2\theta \dot{\varphi}}{\sqrt{1-\cos^2\theta \dot{\varphi}^2-\dot{\theta}^2}}+N_{\textrm{D3}}^{\textrm{D5}}\sin\theta. 
\end{align}
%EOM
The Euler-Lagrange equation for $\theta$ is
\begin{align}
0&=
\frac{d}{dt} \frac{\partial L}{\partial \dot{\theta}}
-\frac{\partial L}{\partial \theta}
\nonumber\\
&=
\frac{64\tilde{c}\tilde{d}\xi\ddot{\theta}}{\sqrt{1-\cos^2\theta \dot{\varphi}^2-\dot{\theta}^2}}
-\frac{32\tilde{c}\tilde{d}\xi\dot{\theta}(2\sin\theta\cos\theta\ \dot{\theta}\dot{\varphi}^2-2\dot{\theta}\ddot{\theta}-2\cos^2\theta\ \dot{\varphi}\ddot{\varphi})}
{(1-\cos^2\theta \dot{\varphi}^2-\dot{\theta})^{\frac32}}
\nonumber\\
&
+\cos\theta \dot{\varphi}
\left[
-N_{\textrm{D3}}^{\textrm{D5}}+\frac{64\tilde{c}\tilde{d}\xi\sin\theta \dot{\varphi}}
{\sqrt{1-\cos^2\theta \dot{\varphi}^2-\dot{\theta}^2}}
\right]. 
\end{align}
We find a giant graviton solution
\begin{align}
\label{GG_sol}
\theta&=\textrm{const.},&
\dot{\varphi}&=1, &
(64\tilde{c}\tilde{d}\xi)^2&={N_{\textrm{D3}}^{\textrm{D5}}}^2. 
\end{align}

%elimination of \varphi => Routhian
Let us consider the giant graviton solution (\ref{GG_sol}) with $\theta$ being constant.  
Since the Lagrangian (\ref{Lag_ETWGG1/2}) does not involve the variable $\varphi$,  
it is a cyclic coordinate and its canonical momentum $P_{\varphi}$ is the integral of motion 
so that it is fixed to be a constant $P$. 
One can eliminate cyclic coordinate from the Lagrangian by constructing a new Lagrangian, which is called the Routhian 
\begin{align}
\label{Routhian0}
R&=L-\frac{\partial L}{\partial \dot{\varphi}}\dot{\varphi}. 
\end{align}
Then the Hamiltonian is simply given by the negative value of the Routhian. 
We obtain
\begin{align}
H&=\sqrt{{N_{\textrm{D3}}^{\textrm{D5}}}^2+\frac{(P-N_{\textrm{D3}}^{\textrm{D5}}\sin\theta)^2}{\cos^2\theta}}. 
\end{align}
For fixed $P$ the Hamiltonian can be viewed as the potential 
that determines the equilibrium configuration of the spherical D3-brane.  
Let us find the local minimum of the Hamiltonian at $\xi =\xi_{0},\theta =\theta_{0}$. 
We find 
\begin{align}
\frac{\partial H}{\partial \theta}
&=
\frac{({N_{\textrm{D3}}^{\textrm{D5}}}\sin\theta-P)({N_{\textrm{D3}}^{\textrm{D5}}}-P\sin\theta)}
{\cos^2\theta \sqrt{{N_{\textrm{D3}}^{\textrm{D5}}}^2-2{N_{\textrm{D3}}^{\textrm{D5}}}P\sin\theta+P^2}}. 
\end{align}
The minimum can appear at $\theta=\theta_0$ where the first derivative with respect to $\theta$ is zero. 
%As the solution should contain the trivial minimum with $\theta_0=0$, 
We find the solution
\begin{align}
\label{theta0_sol}
\sin\theta_0&=\frac{P}{N_{\textrm{D3}}^{\textrm{D5}}}. 
\end{align}
The maximal giant appears when $\theta=\frac{\pi}{2}$ 
so that its angular momentum is equal to the maximal value $N_{\textrm{D3}}^{\textrm{D5}}$. 
At the non-trivial minimum we get the energy
\begin{align}
\label{giant_ene}
E&=H_{\textrm{min}}=N_{\textrm{D3}}^{\textrm{D5}}. 
\end{align}
It follows from (\ref{giant_ene}) that 
the energy of giant graviton in the ETW brane with a single stack of $5$-branes is $N_{\textrm{D3}}^{\textrm{D5}}$ 
and that these states satisfy the BPS bound 
as their energies are saturated by their angular momentum  
corresponding to the $U(1)$ charge, the Cartan generator of the $SU(2)_C$ R-symmetry. 
Therefore these states contribute to the Coulomb limit of the index. 

To summarize, the spectrum of energy and angular momentum of the giant graviton in the ETW brane geometries 
with two asymptotic $AdS_5\times S^5$ regions are bounded by $N_{\textrm{D3}}^{\textrm{D5}}$. 
In particular, when there is a single D5$'$-brane, i.e. $N_{\textrm{D5}}=1$, 
it is simply bounded by $(N-M)$ according to (\ref{D3ch_D5a}).

%%%%%%%%%%%%%%%%%%%%%%%%%%%%%%%%%%%%%%%%%%%%%%%
\subsubsection{Quadratic fluctuations}
%%%%%%%%%%%%%%%%%%%%%%%%%%%%%%%%%%%%%%%%%%%%%%%
Consider the bosonic fluctuations around maximal giant graviton solution
\begin{align}
\theta(t)&=\frac{\pi}{2}+\delta \theta(t),& 
\varphi(t)&=t+\delta\phi(t). 
\end{align}
We expand the Lagrangian (\ref{Lag_ETWGG1/2}) around the giant graviton solution. Our methods here are similar to those of \cite{Eleftheriou:2023jxr,Das:2000st}. 
Instead of $\theta$ and $\varphi$ we introduce the coordinates 
\begin{align}
\sigma&=\frac{\pi}{2}-\theta, & 
\phi&=\varphi-t.  
\end{align}

Then we get the following Lagrangian for the quadratic fluctuations 
\begin{align}
L_{\textrm{fluctuation}}&=N_{\textrm{D3}}^{\textrm{D5}} L^{(2)}_{\textrm{fluctuation}}
\end{align}
where 
\begin{align}
L^{(2)}_{\textrm{fluctuation}}&=
\frac12 \sigma^2 \dot{\phi}^2
+\frac12 \dot{\sigma}^2
-\frac12 \sigma^2\dot{\phi}
+\dot{\phi}. 
\end{align}
Since $\sin \theta$ $\approx$ $1-\frac12\sigma^2$ and $\cos^2\theta$ $\approx$ $\sigma^2$ near $\theta=\frac{\pi}{2}$, 
we have
\begin{align}
\label{app_lag1}
-\sqrt{1-\cos^2\theta \dot{\varphi}^2-\dot{\theta}^2}
&\approx 
-1+\frac12 \sigma^2 \dot{\phi}^2+\frac12 \dot{\sigma}^2, \\
\label{app_lag2}
\sin\theta \dot{\phi}
&\approx -\frac12 \sigma^2\dot{\phi}+\dot{\phi}. 
\end{align}
By making use of (\ref{app_lag1}) and (\ref{app_lag2}) and changing variables $(\sigma,\phi)$ to $(x,y)$ defined by 
\begin{align}
\sigma^2&=x^2+y^2,& 
\frac{y}{x}&=\tan \phi, 
\end{align}
we obtain 
\begin{align}
L^{(2)}_{\textrm{fluctuation}}
&=\frac12 (\dot{x}^2+\dot{y}^2)
+\frac{1}{2}(x\dot{y}-\dot{x}y)-\frac{x\dot{y}-\dot{x}y}{x^2+y^2}. 
\end{align}
Thus we have
\begin{align}
\label{1gg_lag}
L_{\textrm{fluctuation}}
&=\frac{N_{\textrm{D3}}^{\textrm{D5}}}{2} (\dot{x}^2+\dot{y}^2)
+\frac{N_{\textrm{D3}}^{\textrm{D5}}}{2}(x\dot{y}-\dot{x}y)
-\frac{N_{\textrm{D3}}^{\textrm{D5}}(x\dot{y}-\dot{x}y)}{x^2+y^2}. 
\end{align}
The Lagrangian (\ref{1gg_lag}) takes the same form as the Lagrangian for 
quadratic fluctuations of a giant graviton in the $AdS_5$ $\times$ $S^5$ derived in \cite{Eleftheriou:2023jxr}. 
\footnote{Also see \cite{Lee:2023iil,Deddo:2024liu} for the holographic calculations of the half-BPS giant graviton index.}
Therefore the spectrum of the half-BPS maximal giant in the ETW brane is expected to have the same structure as those in the $AdS_5\times S^5$.  
The first two terms can be viewed as the Lagrangian for the Landau problem 
describing a particle on a two-dimensional plane with a magnetic flux. 
As discussed in \cite{Eleftheriou:2023jxr}, the last term effectively induces a shift of $N_{\textrm{D3}}^{\textrm{D5}}$ units of the angular momentum. 
Accordingly, the index of $k$ giants in the ETW brane region will be given by 
\footnote{According to our convention of the half-BPS limit defined by (\ref{H_lim}) and (\ref{C_lim}), the half-BPS indices are Taylor series in $\mathfrak{q}^2$ rather than $\mathfrak{q}$. }
\begin{align}
\label{1/2gg_index}
\frac{(-1)^k \mathfrak{q}^{k(k+1)}}
{(\mathfrak{q}^2;\mathfrak{q}^2)_{k}}
\mathfrak{q}^{2kN_{\textrm{D3}}^{\textrm{D5}}}. 
\end{align}
In the next section, e.g. (\ref{gg_Neu_Hexp}), 
we will see from the gauge theory analysis that the index (\ref{1/2gg_index}) is obtained from the giant graviton expansion of the half-BPS limit of the half-index 
when $N_{\textrm{D3}}^{\textrm{D5}}=N$, with $M=0$, corresponding to the $AdS_4$ ETW brane geometry with a single asymptotic $AdS_5\times S^5$ region. 

However, there is a caveat when $N_{\textrm{D3}}^{\textrm{D5}}=N-M$,  with $M\neq 0$, 
where the $AdS_4$ bagpipe geometry with two asymptotic $AdS_5\times S^5$ regions 
can have extra giants in the $AdS_5\times S^5$ regions with $M$ units of $5$-form flux 
as there are $M$ D3-branes that do not terminate on the $5$-brane. 
Related to this, the spectrum of the fluctuations of giants should be understood by 
taking into account the two types of giants. 
In fact, as will see, the giant graviton expansion of the half-BPS limit of the half-index indicates 
that the spectrum of the fluctuation modes contains such contributions. 

There exist the $1/8$-BPS giants in the ETW brane region. 
They are spherically symmetric on the three-cycle $C_{(i)}$ formed by $S^2_{(i)}$ fibration over a segment on the boundary of the Riemann surface. 
The $1/8$-BPS giants in the bagpipe geometry will have a similar structure of the spectrum 
as we will find that the half-indices admit the giant graviton expansion with respect to $x^{kN_{\textrm{D3}}^{\textrm{D5}}}$ and $x^{mM}$, which encode the spectra of the $1/8$-BPS giants of wrapping number $k$ in the ETW brane region and those of wrapping number $m$ in the asymptotic $AdS_5\times S^5$ region. 
It would be an interesting future direction to explore these $1/8$-BPS giant graviton configurations in the gravity side and their cousins. 
%%%%%%%%%%%%%%%%%%%%%%%%%%%%%%%%%%%%%%%%%%%%%%%
%%%%%%%%%%%%%%%%%%%%%%%%%%%%%%%%%%%%%%%%%%%%%%%
\section{Half-indices of $\mathcal{N}=4$ $U(N)$ SYM theory}
\label{sec_hindex}
%%%%%%%%%%%%%%%%%%%%%%%%%%%%%%%%%%%%%%%%%%%%%%%
%%%%%%%%%%%%%%%%%%%%%%%%%%%%%%%%%%%%%%%%%%%%%%%
The spectrum of the BPS local operators which obey the half-BPS boundary conditions $\mathcal{B}$
in $\mathcal{N}=4$ SYM theory of gauge group $G$ can be evaluated from the half-index \cite{Gaiotto:2019jvo} 
\footnote{Also see \cite{Dimofte:2011py,Gang:2012yr}.}
\begin{align}
\label{hindex_def}
\mathbb{II}_{\mathcal{B}}^{\textrm{4d $G$}}(t,z;q)
&={\Tr}_{\mathcal{H}}(-1)^{F} q^{J+\frac{H+C}{4}}t^{H-C}z^f, 
\end{align}
where $F$ is the Fermion number operator, $J$ the spin, 
$H,C$ the Cartan generators of the $SU(2)_H$ and $SU(2)_C$ factors of the R-symmetry group 
and $f$ the Cartan generators of the other global symmetries. 
The trace is taken over the cohomology $\mathcal{H}$ of the chosen supercharges 
which is a part of the 3d $\mathcal{N}=4$ superalgebra preserved by the half-BPS boundary conditions. 
In general, the half-index (\ref{hindex_def}) is a formal power series in $q^{1/4}$ and a Laurent polynomial in $t$. 

As the half-index can be viewed as a generalized supersymmetric index of 3d $\mathcal{N}=4$ full index which contains additional 4d degrees freedom, 
it admits two limits, the Higgs limit and Coulomb limit \cite{Razamat:2014pta}
\begin{align}
\label{H_lim}
{\mathbb{II}_{\mathcal{B}}^{\textrm{4d $G$}}}^{(H)}(z;\mathfrak{q})
&=\lim_{
\begin{smallmatrix}
\textrm{$\mathfrak{q}:=q^{1/4}t$: fixed}\\
q\rightarrow 0\\
\end{smallmatrix}
}
\mathbb{II}_{\mathcal{B}}^{\textrm{4d $G$}}(t,z;q),\\
\label{C_lim}
{\mathbb{II}_{\mathcal{B}}^{\textrm{4d $G$}}}^{(C)}(z;\mathfrak{q})
&=\lim_{
\begin{smallmatrix}
\textrm{$\mathfrak{q}:=q^{1/4}t^{-1}$: fixed}\\
q\rightarrow 0\\
\end{smallmatrix}
}
\mathbb{II}_{\mathcal{B}}^{\textrm{4d $G$}}(t,z;q). 
\end{align}
In the Higgs (resp. Coulomb) limit, 
the indices count the half-BPS local operators parameterizing the Higgs branch (resp. Coulomb branch). 
The fugacity $\mathfrak{q}$ is coupled to the Cartan generator $H$ (resp. $C$) of $SU(2)_H$ (resp. $SU(2)_C$) R-symmetry group. 
In the gravity dual geometries, $H$ and $C$ correspond to 
the $U(1)$ angular momentum in $S^2_{(2)}$ and that in $S^2_{(1)}$ carried by the giant gravitons in the ETW brane. 

When the boundary condition $\mathcal{B}$ involves additional boundary degrees of freedom, 
the extra boundary degrees of freedom associated with $\mathcal{B}$ can be obtained from the normalized half-index as a quotient 
\begin{align}
\label{norm_hindex}
\mathcal{II}_{\mathcal{B}}^{U(N)}
&=
\frac{\mathbb{II}_{\mathcal{B}}^{U(N)}}
{\mathcal{I}^{U(N)}}, 
\end{align}
where $\mathcal{I}^{U(N)}$ is the Schur index of $\mathcal{N}=4$ $U(N)$ SYM theory. 

In order to express the indices we introduce the $q$-shifted factorial defined by
\begin{align}
\label{qpoch_def}
(a;q)_{0}&:=1,\qquad
(a;q)_{n}:=\prod_{k=0}^{n-1}(1-aq^{k}),\qquad 
(q)_{n}:=\prod_{k=1}^{n}(1-q^{k}),
\nonumber \\
(a;q)_{\infty}&:=\prod_{k=0}^{\infty}(1-aq^{k}),\qquad 
(q)_{\infty}:=\prod_{k=1}^{\infty} (1-q^k), 
\end{align}
where $a$ and $q$ are complex numbers with $|q|<1$. 
For simplicity we use the following notation: 
\begin{align}
(x^{\pm};q)_{n}:=(x;q)_{n}(x^{-1};q)_{n}. 
\end{align}

%%%%%%%%%%%%%%%%%%%%%%%%%%%%%%%%%%%%%%%%%%%%%%%
\subsection{Neumann and Nahm pole boundary conditions}
%%%%%%%%%%%%%%%%%%%%%%%%%%%%%%%%%%%%%%%%%%%%%%%
%\begin{comment}
\begin{figure}
\centering
\scalebox{0.8}{
\begin{tikzpicture}
\filldraw [fill=yellow!20!white,draw=white] (-6,0) -- (-2,0) -- (-2,4) -- (-6,4);
\draw (-6,0) -- (-6,4); 
\node at (-4,2) {$N$ D3};
\node at (-6,4.5) {NS5};
\filldraw [fill=yellow!20!white,draw=white] (2,0) -- (6,0) -- (6,4) -- (2,4);
\draw (2,0)[dashed] -- (2,4); 
\node at (4,2) {$N$ D3};
\node at (2,4.5) {D5$'$}; 
\end{tikzpicture}
}
\caption{The Neumann boundary condition for 4d $\mathcal{N}=4$ $U(N)$ SYM theory realized by a single NS5-brane (left). 
The regular Nahm pole boundary condition for 4d $\mathcal{N}=4$ $U(N)$ SYM theory realized by a single D5$'$-brane (right). 
The horizontal direction is $x^6$. }
\label{fig_NeuNahm}
\end{figure}
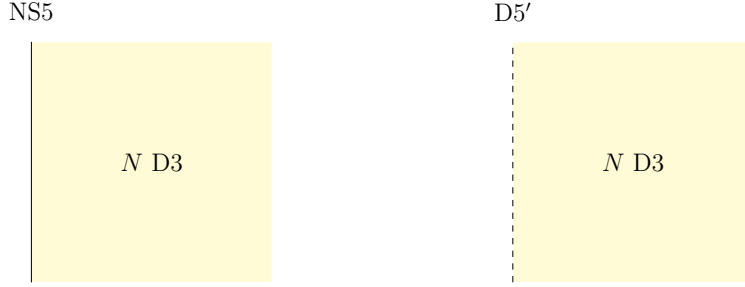
%\end{comment}

Let us consider the Neumann boundary condition in $\mathcal{N}=4$ $U(N)$ SYM theory 
which is realized by a single NS5-brane on which D3-branes terminate 
and the Nahm pole boundary condition in $\mathcal{N}=4$ $U(N)$ SYM theory 
as its S-dual configuration constructed by a single D5$'$-brane (see Figure \ref{fig_NeuNahm}). 
\footnote{
The Neumann boundary condition $\mathcal{N}$ preserving the 3d $\mathcal{N}=4$ vector multiplet 
is constructed by a single NS5-brane with world-volume in the $012345$ direction. 
Note that this is different from the Neumann boundary condition $\mathcal{N}'$ realized by the NS5$'$-brane in (\ref{branesetup}) 
preserving the 3d $\mathcal{N}=4$ twisted verctor multiplet.  
}

The gravity dual geometries can be obtained from the genus $g=1$ solutions   
by further taking the limit where the $5$-form flux at $e_1=0$ vanishes while that at $e_4=\infty$ is kept finite. 
The limit corresponding to a single stack of D5$'$-branes can be realized from the genus $g=1$ solution (\ref{collapse_2}) 
by taking 
\footnote{Here we focus on the D5$'$-brane configuration. See \cite{Aharony:2011yc} for the NS5-brane configuration. }
\begin{align}
\beta&\rightarrow 0,&
\alpha_1&\rightarrow 0, 
\end{align}
while keeping $l^2$ finite. 
This limit can be viewed as a new kind of degeneration limit of a subset $(\alpha e \beta)$. 
In this limit we have
\begin{align}
\Delta^1&=0, \\
\Delta^4&=-l^2-\alpha_2. 
\end{align}
and 
\begin{align}
\alpha_2<-l^2<0. 
\end{align}
Consequently, the asymptotic $AdS_5\times S^5$ region at $u=0$ is closed off 
and the asymptotic $AdS_5\times S^5$ at infinity is left over. 
The differentials become
\begin{align}
\partial h&=-\frac{i}{\sqrt{u}}\frac{u-\alpha_2}{u+l^2}du, \\
\partial \hat{h}&=-\frac{1}{\sqrt{u}}du. 
\end{align}
The harmonic functions are 
\begin{align}
h&=-2i (w-\bar{w})-\frac{l^2+\alpha_2}{l}\ln \frac{|w+il|^2}{|w-il|^2}, \\
\hat{h}&=-2(w+\bar{w}). 
\end{align}

Let us focus on the case with a single D5$'$-brane. 
Since $N_{\textrm{D5}}=1$ we have 
\begin{align}
\alpha_2=-l^2+\frac{l}{16\pi^2}. 
\end{align}
Hence one finds that 
\begin{align}
N=N_{\textrm{D3}_4}&=32\pi l. 
\end{align}

Let us now study the spectrum of the excitations by turning to the analysis in the dual gauge theory. 
The half-index of the Neumann boundary condition $\mathcal{N}$ for $\mathcal{N}=4$ $U(N)$ SYM theory is given by \cite{Gaiotto:2019jvo}
\begin{align}
\label{un_Neu}
\mathbb{II}_{\mathcal{N}}^{\textrm{4d $U(N)$}}
&=
\frac{1}{N!}
\frac{(q)_{\infty}^N}{(q^{\frac12}t^{-2};q)_{\infty}^N}
\oint \prod_{i=1}^{N}
\frac{ds_i}{2\pi is_i}
\prod_{i\neq j}
\frac{(\frac{s_i}{s_j};q)_{\infty}}{(q^{\frac12}t^{-2} \frac{s_i}{s_j};q)_{\infty}}. 
\end{align}

The Neumann boundary condition for $U(N)$ gauge theory is S-dual to the regular Nahm pole boundary condition \cite{Gaiotto:2008ak}. 
The regular Nahm pole boundary condition is realized by a single D5$'$-brane on which D3-branes end. 
The Nahm pole half-index can be obtained from the Dirichlet half-index by means of the Higgsing procedure \cite{Gaiotto:2012xa}. 
It is calculated as \cite{Gaiotto:2019jvo}
\begin{align}
\label{un_Nahm}
\mathbb{II}_{\textrm{Nahm}'}^{\textrm{4d $U(N)$}}
&=\prod_{k=1}^{N}
\frac{(q^{\frac{k+1}{2}}t^{-2(k-1)};q)_{\infty}}{(q^{\frac{k}{2}}t^{-2k};q)_{\infty}}
\nonumber\\
&=\frac{(q)_{\infty}}{(q^{\frac12}t^{-2};q^{\frac12}t^{-2})_{N-1} (q^{\frac{N}{2}} t^{-2N};q)_{\infty}}. 
\end{align}
In fact, the two half-indices (\ref{un_Neu}) and (\ref{un_Nahm}) are equivalent \cite{Gaiotto:2019jvo}
\begin{align}
\mathbb{II}_{\mathcal{N}}^{\textrm{4d $U(N)$}}
&=\mathbb{II}_{\textrm{Nahm}'}^{\textrm{4d $U(N)$}}. 
\end{align}

In the Coulomb limit or the half-BPS limit, we get
\begin{align}
\label{un_Neu_C}
{\mathbb{II}_{\mathcal{N}}^{\textrm{4d $U(N)$}}}^{(C)}
&=
{\mathbb{II}_{\textrm{Nahm}'}^{\textrm{4d $U(N)$}}}^{(C)}
=
\frac{1}{(\mathfrak{q}^2;\mathfrak{q}^2)_N}. 
\end{align}
We note that since the corresponding boundary conditions contain no additional boundary degrees of freedom, 
this simply agrees with the half-BPS limit of the full index. 

%%%%%%%%%%%%%%%%%%%%%%%%%%%%%%%%%%%%%%%%%%%%%%%
\subsubsection{Large $N$ limit}
%%%%%%%%%%%%%%%%%%%%%%%%%%%%%%%%%%%%%%%%%%%%%%%
As mentioned above, the dual geometry of Type IIB supergravity is realized as a genus $g=1$ solution 
with a single asymptotic $AdS_5\times S^5$ region at $u=\infty$. 
The $AdS_4$ vacuum associated with the compact internal $6$-manifold 
in the ETW brane region has low-lying fluctuation fields as well as KK towers of massive modes. 

The spectrum of such fluctuation fields is expected to be obtained from the gauge theory side 
by taking the large $N$ limit of the half-indices. 
In the large $N$ limit, the Neumann half-index (\ref{un_Neu}) 
or equivalently the Nahm pole half-index (\ref{un_Nahm}) becomes 
\begin{align}
\label{largeN_Neu}
\mathbb{II}_{\mathcal{N}}^{\textrm{4d $U(\infty)$}}
&=\mathbb{II}_{\textrm{Nahm}'}^{\textrm{4d $U(\infty)$}}
=\prod_{n=1}^{\infty}
\frac{1-q^n}{1-q^{\frac{n}{2}}t^{-2n}}. 
\end{align}
By taking the plethystic logarithm \cite{MR1601666}, 
we obtain the single particle gravity index 
\begin{align}
\label{ETW_index1}
i^{\textrm{ETW}}&=-\frac{q}{1-q}+\frac{q^{\frac12}t^{-2}}{1-q^{\frac12}t^{-2}}. 
\end{align}
This indicates that there is a pair of infinite bosonic and a fermionic KK towers of $1/8$-BPS excitations over the $AdS_4$ ETW brane geometry 
constructed by a single $5$-brane on which semi-infinite D3-branes terminate. 

Let us examine the asymptotic degeneracy of states with large charge. 
In the unflavored limit, cancellation between the bosonic and fermionic excitations 
reduces the large $N$ half-index (\ref{largeN_Neu}) to
\begin{align}
\label{largeN_Neu_u}
\mathbb{II}_{\mathcal{N}}^{\textrm{4d $U(\infty)$}}
&=\mathbb{II}_{\textrm{Nahm}'}^{\textrm{4d $U(\infty)$}}
=\prod_{n=1}^{\infty}(1+q^{\frac{n}{2}})
=\sum_{n\ge 0} d_{\mathcal{N}}^{\textrm{4d $U(\infty)$}}(n)q^{\frac{n}{2}}. 
\end{align}
This is identified with the generating function for the partitions of $n$ into distinct parts. 
Applying the Meinardus theorem \cite{MR62781, MR1634067}, 
one finds the asymptotic degeneracy of the boundary local operators
\begin{align}
\label{d_Neu}
d_{\mathcal{N}}^{\textrm{4d $U(\infty)$}}(n)
&\sim \frac{1}{4\cdot 3^{\frac14}n^{\frac34}}
\exp\left[
\frac{\pi}{3^{\frac12}}n^{\frac12}
\right]. 
\end{align}
The asymptotic degeneracy of the bulk local operators obtained from the large $N$ full-index
\begin{align}
\mathcal{I}^{U(\infty)}
&=\prod_{n=1}^{\infty}
\frac{1+q^{\frac{n}{2}}}{1-q^{\frac{n}{2}}}
=\sum_{n\ge0}
d^{\textrm{4d $U(\infty)$}}(n)q^{\frac{n}{2}}
\end{align}
is given by \cite{Hatsuda:2022xdv}
\begin{align}
d^{\textrm{4d $U(\infty)$}}(n)&\sim 
\frac{1}{8n}
\exp\left[
\pi n^{\frac12}
\right], 
\end{align}
which is greater than the asymptotic degeneracy (\ref{d_Neu}) of the boundary states. 

The Coulomb limit of the large $N$ half-index (\ref{largeN_Neu}) is
\begin{align}
{\mathbb{II}_{\mathcal{N}}^{\textrm{4d $U(\infty)$}}}^{(C)}
&=
{\mathbb{II}_{\textrm{Nahm}'}^{\textrm{4d $U(\infty)$}}}^{(C)}
=\frac{1}{(\mathfrak{q}^2;\mathfrak{q}^2)_{\infty}}. 
\end{align}

%%%%%%%%%%%%%%%%%%%%%%%%%%%%%%%%%%%%%%%%%%%%%%%
\subsubsection{Giant graviton expansion}
%%%%%%%%%%%%%%%%%%%%%%%%%%%%%%%%%%%%%%%%%%%%%%%
In this case, the gravity dual ETW brane geometry supports giant gravitons with angular momentum $N_{\textrm{D3}}^{\textrm{D5}}$ $=$ $N$ 
wrapping the three-cycle as the fibration of the two-sphere over the segment on the boundary of Riemann surface. 
Thus the finite $N$ correction to the half-index is expected to be controlled by giant gravitons in the ETW brane. 
In the Coulomb limit, the half-index has the following expansion: 
\begin{align}
\label{gg_Neu_Hexp}
\frac{
{\mathbb{II}_{\mathcal{N}}^{\textrm{4d $U(N)$}}}^{(C)}
}
{
{\mathbb{II}_{\mathcal{N}}^{\textrm{4d $U(\infty)$}}}^{(C)}
}&=\sum_{k=0}^{\infty}\mathfrak{q}^{2Nk}\hat{F}_{k}^{(C)}(\mathfrak{q}), 
\end{align}
where 
\begin{align}
\label{gg_Neu_H}
\hat{F}_{k}^{(C)}(\mathfrak{q})&=
\frac{(-1)^k \mathfrak{q}^{k(k+1)}}{(\mathfrak{q}^2;\mathfrak{q}^2)_{k}}. 
\end{align}
The term $\mathfrak{q}^{2Nk}\hat{F}_{k}^{(C)}(\mathfrak{q})$ in the expansion (\ref{gg_Neu_Hexp}) 
is exactly same as the giant graviton index (\ref{1/2gg_index}) which is obtained from the semiclassical analysis of the wrapped D3-branes.  
It encodes the spectrum of $k$ half-BPS giant gravitons as wrapped D3-branes in the ETW brane. 

Under $\mathfrak{q}\rightarrow \mathfrak{q}^{-1}$, 
the giant graviton index (\ref{gg_Neu_H}) becomes the Coulomb limit (\ref{un_Neu_C}) of the half-index
\begin{align}
\hat{F}_{k}^{(C)}(\mathfrak{q}^{-1})&=
{\mathbb{II}_{\mathcal{N}}^{\textrm{4d $U(k)$}}}^{(C)}(\mathfrak{q}). 
\end{align}
Thus the giant graviton expansion for the half-BPS giants can be rewritten as
\begin{align}
\frac{
{\mathbb{II}_{\mathcal{N}}^{\textrm{4d $U(N)$}}}^{(C)}
(\mathfrak{q})
}
{
{\mathbb{II}_{\mathcal{N}}^{\textrm{4d $U(\infty)$}}}^{(C)}
(\mathfrak{q})
}&=\sum_{k=0}^{\infty}\mathfrak{q}^{2Nk} 
{\mathbb{II}_{\mathcal{N}}^{\textrm{4d $U(k)$}}}^{(C)}(\mathfrak{q}^{-1}). 
\end{align}

Unlike the Coulomb limit corresponding to the half-BPS giant gravitons, 
the story for the half-index itself is rather different from the full-index. 
From the finite $N$ half-index (\ref{largeN_Neu}) and the large $N$ half-index (\ref{un_Nahm}) we obtain
\begin{align}
\frac{\mathbb{II}_{\mathcal{N}}^{\textrm{4d $U(N)$}}}
{\mathbb{II}_{\mathcal{N}}^{\textrm{4d $U(\infty)$}}}
&=\frac{(q^{\frac{N}{2}}t^{-2N};q^{\frac12}t^{-2})_{\infty}}
{(q^{\frac{N}{2}}t^{-2N};q)_{\infty}}. 
\end{align}
Let us use new variables
\begin{align}
\label{xy_var}
x=q^{\frac12}t^{-2},\qquad y=q^{\frac12}t^{2},\qquad q=xy.
\end{align}
Then we have 
\begin{align}
\frac{\mathbb{II}_{\mathcal{N}}^{\textrm{4d $U(N)$}}}{\mathbb{II}_{\mathcal{N}}^{\textrm{4d $U(\infty)$}}}
&=\frac{(x^N;x)_{\infty}}{(x^N;q)_{\infty}}.
\end{align}
Using the identities
\begin{align}
(z;q)_\infty&=\sum_{k=0}^\infty z^{k}\frac{(-1)^k q^{\frac{k(k-1)}{2}}}{(q;q)_k}, \\
\frac{1}{(z;q)_\infty}&=\sum_{k=0}^\infty z^{k}\frac{1}{(q;q)_k},
\end{align}
we can easily find the simple-sum giant graviton expansion of the half-index:
\begin{align}
\frac{\mathbb{II}_{\mathcal{N}}^{\textrm{4d $U(N)$}}}{\mathbb{II}_{\mathcal{N}}^{\textrm{4d $U(\infty)$}}}
=\frac{(x^N;x)_{\infty}}{(x^N;q)_{\infty}}=\sum_{k=0}^\infty x^{kN} \hat{F}_k(x;y),
\end{align}
where
\begin{align}
\label{GGhindex_Neu0}
\hat{F}_k(x;y):=\sum_{\ell=0}^k \frac{(-1)^\ell x^{\frac{\ell(\ell-1)}{2}}}{(x;x)_\ell (q; q)_{k-\ell}}.
\end{align}

Since the actions of the symmetry generators for the effective theory on the giant gravitons are different, 
the giant graviton index can be identified by redefining fugacities \cite{Arai:2019xmp,Gaiotto:2021xce}
\begin{align}
\label{x_change}
\sigma_x: (x,y,q)&\rightarrow (x^{-1},q,y),\\
\label{y_change}
\sigma_y: (x,y,q)&\rightarrow (q,y^{-1},x). 
\end{align}
Using
\begin{align}
(x;x)_\ell=(-1)^\ell x^{\frac{\ell(\ell+1)}{2}}(x^{-1};x^{-1})_\ell,
\end{align}
we can rewrite (\ref{GGhindex_Neu0}) as
\begin{align}
\hat{F}_k(x;y)=\sum_{\ell=0}^k \frac{x^{-\ell}}{(x^{-1};x^{-1})_\ell (q; q)_{k-\ell}}=\sigma_x [F_k(x;y)],
\end{align}
where
\begin{align}
\label{GGhindex_Neu}
F_k(x;y):=\sum_{\ell=0}^k \frac{x^\ell}{(x;x)_\ell (y;y)_{k-\ell}}
\end{align}
is the giant graviton index. 
Interestingly, the giant graviton index (\ref{GGhindex_Neu}) 
turns out to be identical to the large $N$ normalized two-point function of the Wilson lines in the rank-$k$ (anti)symmetric representation 
(see (5.16)-(5.18) in \cite{Hatsuda:2023iwi} or (3.29) in \cite{Hatsuda:2023imp}) !
We conclude
\begin{align}
F_k(x;y)&=\langle \mathcal{W}_{(k)}\mathcal{W}_{\overline{(k)}} \rangle^{U(\infty)}
=\langle \mathcal{W}_{(1^k)}\mathcal{W}_{\overline{(1^k)}} \rangle^{U(\infty)}. 
\end{align}
In other words, 
the $1/8$-BPS index of $k$ giant gravitons in the $AdS_4$ ETW brane is dual to 
the large $N$ normalized two-point function which describes 
the spectrum of a probe D3-brane giant (or D5-brane giant) wrapping $AdS_2$ with $k$ units of electric flux 
corresponding to $k$ fundamental strings. 

In summary, we have
\begin{align}
\frac{\mathbb{II}_{\mathcal{N}}^{\textrm{4d $U(N)$}}}{\mathbb{II}_{\mathcal{N}}^{\textrm{4d $U(\infty)$}}}
=\sum_{k=0}^\infty x^{kN} \sigma_x [\langle \mathcal{W}_{(1^k)}\mathcal{W}_{\overline{(1^k)}} \rangle^{U(\infty)}].
\end{align}

Very recently, it is numerically found and conjectured in \cite{Imamura:2024pgp} (also see \cite{Beccaria:2024oif} for related analysis) 
that the finite $k$ corrections to the large $N$ two-point function $F_k(x,y)$ is conversely related to the half-index:
\begin{align}
\frac{F_k(x;y)}{F_\infty(x;y)}&=\sum_{m=0}^\infty x^{km} \mathcal{F}_{m,0},
\label{eq:brane-expansion}
\end{align}
where
\begin{align}
\mathcal{F}_{m,0}=\sigma_x \mathbb{II}_{\mathcal{N}}^{\textrm{4d $U(m)$}}.
\end{align}
We can show this equality analytically. To do so, we first rewrite $\mathcal{F}_{m,0}$ as
\begin{align}
\label{inv_F}
\mathcal{F}_{m,0}&=\frac{(y;y)_\infty (x^{-m};x^{-1})_\infty}{(x^{-1};x^{-1})_\infty (x^{-m};y)_\infty}
\nonumber\\
&=\frac{(-1)^{m-1}x^{\frac{m(m-1)}{2}}}{(x;x)_{m-1}}\frac{(y;y)_\infty}{(x^{-m};y)_\infty} \nonumber \\
&=(y;y)_\infty \frac{(-1)^{m-1}x^{\frac{m(m-1)}{2}}}{(x;x)_{m-1}}\sum_{\ell=0}^\infty \frac{x^{-m \ell}}{(y;y)_\ell}.
\end{align}
Plugging the expression (\ref{inv_F}) into the right hand side of (\ref{eq:brane-expansion}), we find
\begin{align}
\sum_{m=0}^\infty x^{km} \mathcal{F}_{m,0}&
=(y;y)_\infty \sum_{\ell=0}^\infty \frac{1}{(y;y)_\ell}\sum_{m=0}^\infty \frac{(-1)^{m-1}x^{\frac{m(m-1)}{2}}}{(x;x)_{m-1}}x^{m(k-\ell)} \nonumber \\
%&=(y;y)_\infty \sum_{\ell=0}^\infty \frac{1}{(y;y)_\ell}x^{k-\ell}\sum_{m=0}^\infty \frac{(-1)^{m}x^{\frac{m(m-1)}{2}}}{(x;x)_{m}}x^{m(k-\ell+1)} \nonumber \\
%&=(y;y)_\infty \sum_{\ell=0}^\infty \frac{1}{(y;y)_\ell}x^{k-\ell}(x^{k-\ell+1};x)_\infty \nonumber \\
&=(x;x)_\infty (y;y)_\infty \sum_{\ell=0}^\infty \frac{1}{(y;y)_\ell}\frac{x^{k-\ell}}{(x;x)_{k-\ell}}\nonumber \\
&=(x;x)_\infty (y;y)_\infty \sum_{\ell=0}^k \frac{x^{\ell}}{(x;x)_{\ell}(y;y)_{k-\ell}}.
\end{align}
This is just the ratio $F_k(x;y)/F_\infty(x;y)$.
Therefore, we have just analytically proved that
\begin{align}
\frac{\langle \mathcal{W}_{(1^k)}\mathcal{W}_{\overline{(1^k)}} \rangle^{U(\infty)}}
{\langle \mathcal{W}_{(1^\infty)}\mathcal{W}_{\overline{(1^\infty)}} \rangle^{U(\infty)}}
&=\sum_{m=0}^{\infty} x^{km} 
\sigma_{x}[ \mathbb{II}_{\mathcal{N}}^{\textrm{4d $U(m)$}} ]. 
\end{align}

%%%%%%%%%%%%%%%%%%%%%%%%%%%%%%%%%%%%%%%%%%%%%%%
\subsection{$U(N)|U(M)$ interfaces}
%%%%%%%%%%%%%%%%%%%%%%%%%%%%%%%%%%%%%%%%%%%%%%%
%\begin{comment}
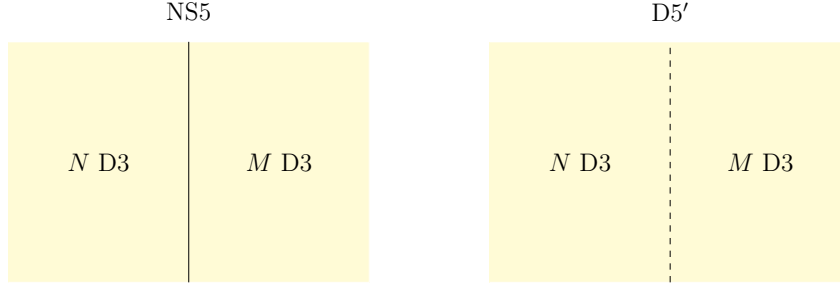
\begin{figure}
\centering
\scalebox{0.8}{
\begin{tikzpicture}
\filldraw [fill=yellow!20!white,draw=white] (-9,0) -- (-3,0) -- (-3,4) -- (-9,4);
\draw (-6,0) -- (-6,4); 
\node at (-7.5,2) {$N$ D3};
\node at (-4.5,2) {$M$ D3};
\node at (-6,4.5) {NS5};
\filldraw [fill=yellow!20!white,draw=white] (-1,0) -- (5,0) -- (5,4) -- (-1,4);
\draw (2,0)[dashed] -- (2,4); 
\node at (0.5,2) {$N$ D3};
\node at (3.5,2) {$M$ D3};
\node at (2,4.5) {D5$'$}; 
\end{tikzpicture}
}
\caption{The 3d interface of NS5-type in 4d $U(N)\times U(M)$ bulk theory (left). 
The dual 3d interface of D5-type (right). }
\label{fig_uN|uM}
\end{figure}
%\end{comment}

%U(N)|U(M)
Next consider the 3d interface constructed by a single NS5-brane between 4d $\mathcal{N}=4$ $U(N)$ and $U(M)$ gauge theories (see Figure \ref{fig_uN|uM}). 
Without loss of generality we assume that $N\ge M$.
It is described by 4d $\mathcal{N}=4$ $U(N)$ SYM theory in the half-space $x^6<0$ 
and 4d $\mathcal{N}=4$ $U(M)$ gauge theory in $x^6>0$. 
At the interface both gauge theories obey the Neumann boundary condition 
and there exists the 3d hypermultiplet transforming in 
the bifundamental representation $(\bf{N},\overline{\bf{M}})$ $\oplus$ $(\overline{\bf{N}},\bf{M})$ of gauge group $U(N)\times U(M)$. 

%S-dual D5 interace
The S-dual 3d interface is constructed by a single D5-brane (see Figure \ref{fig_uN|uM}). 
When $N$ is equal to $M$, the theory is described by 4d $\mathcal{N}=4$ $U(N)$ SYM theory 
coupled to a twisted hypermultiplet transforming in the fundamental representation of $U(N)$ gauge group at the interface. 
When $N$ is greater than $M$, the $U(N)$ gauge group for $x^6<0$ is broken to $U(M)$ 
and the $U(M)$ gauge group is preserved in a whole region at the interface. 
The 4d $U(M)$ vector multiplet further couples to the fundamental 3d chiral multiplets. 

Again the dual $AdS_4$ bagpipe geometries can be found from the genus $g=1$ solutions. 
For the D5$'$-type interface the geometry is obtained from (\ref{collapse_2}), 
which are labeled by four parameters $\alpha_1$, $\alpha_2$, $\beta$ and $l$. 
The solutions have two asymptotic $AdS_5\times S^5$ regions at $u=0$ and $u=\infty$. 

The half-index of the NS5-type $U(N)|U(M)$ interface is given by \cite{Gaiotto:2019jvo}
\begin{align}
\label{un_ns_um_hindex}
\mathbb{II}_{\mathcal{N}}^{U(N)|U(M)}
&=\frac{1}{N!}
\frac{(q)_{\infty}^N}{(q^{\frac12}t^{-2};q)_{\infty}^N}
\oint \prod_{i=1}^N \frac{ds_i}{2\pi is_i}
\prod_{i\neq j}
\frac{\left(\frac{s_i}{s_j};q\right)_{\infty}}
{\left(q^{\frac12}t^{-2}\frac{s_i}{s_j};q \right)_{\infty}}
\nonumber\\
&\times 
\frac{1}{M!}
\frac{(q)_{\infty}^M}{(q^{\frac12}t^{-2};q)_{\infty}^M}
\oint \prod_{i=N+1}^{N+M} \frac{ds_i}{2\pi is_i}
\prod_{i\neq j}
\frac{\left(\frac{s_i}{s_j};q\right)_{\infty}}
{\left(q^{\frac12}t^{-2}\frac{s_i}{s_j};q \right)_{\infty}}
\nonumber\\
&\times 
\prod_{i=1}^{N}\prod_{j=N+1}^{N+M}
\frac{
\left(q^{\frac34}t^{-1}s_i^{\pm}s_{j}^{\mp};q \right)_{\infty}
}
{
\left(q^{\frac14}ts_i^{\pm}s_{j}^{\mp};q \right)_{\infty}
}. 
\end{align}

The half-index of the D5-type $U(N)|U(M)$ interface is given by \cite{Gaiotto:2019jvo}
\begin{align}
\label{un_d5_um_hindex}
\mathbb{II}_{\mathcal{D}'}^{U(N)|U(M)}
&=
\frac{1}{N!}\frac{(q)_{\infty}^{2N}}{(q^{\frac12}t^{\pm2};q)_{\infty}^N}
\oint \prod_{i=1}^{N}
\frac{ds_i}{2\pi is_i}
\prod_{i\neq j}
\frac{
\left(\frac{s_i}{s_j};q\right)_{\infty}
\left(q\frac{s_i}{s_j};q\right)_{\infty}
}
{
\left(q^{\frac12}t^2\frac{s_i}{s_j};q\right)_{\infty}
\left(q^{\frac12}t^{-2}\frac{s_i}{s_j};q\right)_{\infty}
}
\nonumber\\
&\times 
\prod_{k=1}^{N-M}
\frac{(q^{\frac{k+1}{2}}t^{-2(k-1)};q)_{\infty}}
{(q^{\frac{k}{2}}t^{-2k};q)_{\infty}}
\prod_{i=1}^{N}
\frac{(q^{\frac34+\frac{N-M}{4}}t^{1-(N-M)}s_i^{\pm};q)_{\infty}}
{(q^{\frac14+\frac{N-M}{4}}t^{-1-(N-M)}s_i^{\pm};q)_{\infty}}. 
\end{align}

It is numerically confirmed in \cite{Gaiotto:2019jvo} that 
the half-indices (\ref{un_ns_um_hindex}) and (\ref{un_d5_um_hindex}) agree with each other 
as strong evidence of the duality of interfaces. 
Here we propose a new conjectural formula of 
the half-index (\ref{un_ns_um_hindex}) or equivalently (\ref{un_d5_um_hindex}) 
for the general $U(N)|U(M)$ interface constructed by a single $5$-brane.  
We find that the half-indices (\ref{un_ns_um_hindex}) 
and (\ref{un_d5_um_hindex}) can be expressed as 
\begin{align}
\label{un_um_exp}
&
\mathbb{II}_{\mathcal{N}}^{U(N)|U(M)}=\mathbb{II}_{\mathcal{D}'}^{U(N)|U(M)}
\nonumber\\
&=\prod_{k=1}^{N-M}
\frac{(q^{\frac{k+1}{2}}t^{-2(k-1)};q)_{\infty}}
{(q^{\frac{k}{2}}t^{-2k};q)_{\infty}}
\sum_{n_1,\cdots,n_M\ge0}
\frac{(q^{1+n_1};q)_{\infty}(q^{1+\frac{N-M}{2}+n_1}t^{-2(N-M)};q)_{\infty}}
{(q^{\frac12+n_1}t^{-2};q)_{\infty}(q^{\frac12+\frac{N-M}{2}+n_1}t^{-2(N-M+1)};q)_{\infty}}
\nonumber\\
&\times 
\frac{(q^{\frac32+n_1+n_2}t^{-2};q)_{\infty}(q^{\frac32+\frac{N-M}{2}+n_1+n_2}t^{-2(N-M+1)};q)_{\infty}}
{(q^{1+n_1+n_2}t^{-4};q)_{\infty}(q^{1+\frac{N-M}{2}+n_1+n_2}t^{-2(N-M+2)};q)_{\infty}}
\times \cdots
\nonumber\\
&
\times 
\frac{(q^{\frac{M+1}{2}+n_1+\cdots+n_M}t^{-2(M-1)};q)_{\infty}
(q^{\frac{N+1}{2}+n_1+\cdots+n_M}t^{-2(N-1)};q)_{\infty}}
{(q^{\frac{M}{2}+n_1+\cdots+n_M}t^{-2M};q)_{\infty}
(q^{\frac{N}{2}+n_1+\cdots+n_M}t^{-2N};q)_{\infty}}
\nonumber\\
&\times 
(q^{\frac12}t^{2})^{Mn_1+(M-1)n_2+\cdots+n_M}. 
\end{align}
This formula generalizes the results in \cite{Gaiotto:2019jvo}. 
It is factorized into the half-index (\ref{un_Nahm}) for the rank-$(N-M)$ Nahm pole boundary condition and multiple sums 
with $M$ indices. 
When $M=0$, the expression (\ref{un_um_exp}) reproduces the rank-$N$ regular Nahm pole half-index (\ref{un_Nahm}). 

As a special case with $N=M$ we have 
\begin{align}
\label{un_un_exp}
&
\mathbb{II}_{\mathcal{N}}^{U(N)|U(N)}
=\mathbb{II}_{\mathcal{D}'}^{U(N)|U(N)}
\nonumber\\
&=\sum_{n_1,\cdots, n_N\ge0}
\frac{(q^{1+n_1};q)_{\infty}^2 (q^{\frac32+n_1+n_2}t^{-2};q)_{\infty}^2
\cdots (q^{\frac{N+1}{2}+n_1+\cdots+n_N}t^{-2(N-1)};q)_{\infty}^2}
{(q^{\frac12+n_1}t^{-2};q)_{\infty}^2(q^{1+n_1+n_2}t^{-4};q)_{\infty}^2
\cdots(q^{\frac{N}{2}+n_1+\cdots+n_N}t^{-2N};q)_{\infty}^2}
\nonumber\\
&\times 
(q^{\frac12}t^{2})^{Nn_1+(N-1)n_2+\cdots+n_N}. 
\end{align}
Interestingly, we find a similar expression for the $U(N)$ Schur index. 
We conjecture that 
\begin{align}
\label{schur_new}
\mathcal{I}^{U(N)}
&=\sum_{n_1,\cdots, n_N\ge0}
\frac{(q^{1+n_1};q)_{\infty} (q^{\frac32+n_1+n_2}t^{-2};q)_{\infty}
\cdots (q^{\frac{N+1}{2}+n_1+\cdots+n_N}t^{-2(N-1)};q)_{\infty}}
{(q^{\frac12+n_1}t^{-2};q)_{\infty}(q^{1+n_1+n_2}t^{-4};q)_{\infty}
\cdots(q^{\frac{N}{2}+n_1+\cdots+n_N}t^{-2N};q)_{\infty}}
\nonumber\\
&\times 
(q^{\frac12}t^{2})^{Nn_1+(N-1)n_2+\cdots+n_N}. 
\end{align}
Note that the expression (\ref{schur_new}) is obtained by changing the powers of the factors associated with the $q$-factorials in (\ref{un_un_exp}). 
While the exact closed-form expressions for the Schur index is derived in \cite{Hatsuda:2022xdv} by means of the Fermi-gas method, the formula (\ref{schur_new}) takes a different form from them. It would be interesting to give an analytical proof of the equivalence with them. 

%Coulomb limit
Let us consider the Coulomb limit of the $U(N)|U(M)$ interface half-indices.  
In this case, the matrix integral (\ref{un_d5_um_hindex}) can be exactly evaluated according to the inner product of the Hall-Littlewood functions. 
In the Coulomb limit the matrix integral reduces to 
\begin{align}
\label{un_d5_nm_C}
&
{\mathbb{II}_{\mathcal{D}'}^{U(N)|U(M)}}^{(C)}
\nonumber\\
&=
\frac{1}{N!}\frac{1}{(1-\mathfrak{q}^2)^N}
\prod_{k=1}^{N-M}\frac{1}{1-\mathfrak{q}^{2k}}
\oint \prod_{i=1}^{N}
\frac{ds_i}{2\pi is_i}
\prod_{i\neq j}
\frac{1-\frac{s_i}{s_j}}{1-\mathfrak{q}^2 \frac{s_i}{s_j}}
\prod_{i=1}^N 
\frac{1}{1-\mathfrak{q}^{N-M+1}s_i^{\pm}}. 
\end{align}
By means of the orthogonality 
\begin{align}
\frac{1}{N!}
\oint \prod_{i=1}^N \frac{ds_i}{2\pi is_i}
\frac{\prod_{i\neq j} 1-\frac{s_i}{s_j}}{1-\mathfrak{t}\frac{s_i}{s_j}}
P_{\mu}(s;\mathfrak{t})P_{\lambda}(s^{-1};\mathfrak{t})
&=\frac{\delta_{\mu\lambda}}{(\mathfrak{t};\mathfrak{t})_{N-l(\mu)}\prod_{j\ge 1}(\mathfrak{t};\mathfrak{t})_{m_j(\mu)}}
\end{align}
of the Hall-Littlewood functions $P_{\lambda}(s;\mathfrak{t})$, the matrix integral (\ref{un_d5_nm_C}) can be expanded as 
a sum over the Young diagram $\lambda$ whose length $l(\lambda)$ is at most $N$. 
Let us define $n(\lambda)$ $:=$ $\sum_{i\ge} (i-1)\lambda_i$ 
and denote by $|\lambda|$ the number of boxes of $\lambda$ and 
by $m_{j}(\lambda)$ the multiplicity of $\lambda$. 
One finds
\begin{align}
{\mathbb{II}_{\mathcal{D}'}^{U(N)|U(M)}}^{(C)}
&=
\frac{1}{(\mathfrak{q}^2;\mathfrak{q}^2)_{N-M}}
\sum_{\begin{smallmatrix}
\lambda\\
l(\lambda)\le N\\
\end{smallmatrix}}
\frac{\mathfrak{q}^{4n(\lambda)+2|\lambda|(N-M+1)}}
{(\mathfrak{q}^2;\mathfrak{q}^2)_{N-l(\lambda)}
\prod_{j\ge1}(\mathfrak{q}^2;\mathfrak{q}^2)_{m_j(\lambda)}}. 
\end{align}
It is evaluated as
\begin{align}
\label{unum_1/2}
{\mathbb{II}_{\mathcal{D}'}^{U(N)|U(M)}}^{(C)}
&=\frac{1}{(\mathfrak{q}^2;\mathfrak{q}^2)_{N} (\mathfrak{q}^2;\mathfrak{q}^2)_{M}}. 
\end{align}

%%%%%%%%%%%%%%%%%%%%%%%%%%%%%%%%%%%%%%%%%%%%%%%
\subsubsection{Large $N$ limit}
%%%%%%%%%%%%%%%%%%%%%%%%%%%%%%%%%%%%%%%%%%%%%%%
The spectrum of the excitations around the dual $AdS_4$ bagpipe geometry 
with two asymptotic $AdS_5\times S^5$ regions can be obtained 
by taking the large gauge rank limits of the 
$U(N)|U(M)$ interface half-indices (\ref{un_ns_um_hindex}) and (\ref{un_d5_um_hindex}). 

For the $U(N)|U(1)$ interfaces half-indices saturate in the large $N$ limit 
in such a way that the finite $N$ correction shows up at order $q^{\frac{N+1}{2}}$. 
For example, by turning off the flavor fugacity $t$, one finds 
\begin{align}
\mathbb{II}_{\mathcal{N}}^{U(2)|U(1)}
&=1+3q^{1/2}+4q+5q^{3/2}+7q^2+7q^{5/2}+7q^3+11q^{7/2}+\cdots, \\
\mathbb{II}_{\mathcal{N}}^{U(3)|U(1)}
&=1+3q^{1/2}+4q+7q^{3/2}+9q^2+9q^{5/2}+15q^3+15q^{7/2}+\cdots, \\
\mathbb{II}_{\mathcal{N}}^{U(4)|U(1)}
&=1+3q^{1/2}+4q+7q^{3/2}+11q^{2}+11q^{5/2}+17q^3+23q^{7/2}+\cdots, \\
\mathbb{II}_{\mathcal{N}}^{U(5)|U(1)}
&=1+3q^{1/2}+4q+7q^{3/2}+11q^{2}+13q^{5/2}+19q^3+25q^{7/2}+\cdots, \\
\mathbb{II}_{\mathcal{N}}^{U(6)|U(1)}
&=1+3q^{1/2}+4q+7q^{3/2}+11q^{2}+13q^{5/2}+21q^3+27q^{7/2}+\cdots.
\end{align}
In fact, the large $N$ limit of the $U(N)|U(1)$ half-indices is given by
\begin{align}
\label{largeN_u1_hindex}
\mathbb{II}_{\mathcal{N}}^{U(\infty)|U(1)}
&=
\mathbb{II}_{\mathcal{D}'}^{U(\infty)|U(1)}
=
\prod_{n=1}^{\infty}
\frac{1-q^n}{1-q^{\frac{n}{2}}t^{-2n}}
\frac{(q)_{\infty}^2}{(q^{\frac12}t^{2};q)_{\infty} (q^{\frac12}t^{-2};q)_{\infty}}
\end{align}
While the expression (\ref{largeN_u1_hindex}) contains 
the large $N$ index (\ref{largeN_Neu}) that is responsible for the fluctuations of the $AdS_4$ ETW brane, 
it also includes extra factors which coincide with the full-index of $U(1)$ gauge theory. 
The latter capture additional excitations 
due to a D3-brane in the asymptotic $AdS_5\times S^5$ region on the other side of the ETW brane. 

Evaluating the plethystic logrighm of the large $N$ half-index (\ref{largeN_u1_hindex}), 
we obtain the single particle gravity index  
\begin{align}
\label{ETW_1D3_gindex}
i^{\textrm{ETW}+\textrm{$1$ D3}}&=
\underbrace{
\left(
-\frac{q}{1-q}+\frac{q^{\frac12}t^{-2}}{1-q^{\frac12}t^{-2}}
\right)
}_{\textrm{ETW brane}}
+
\underbrace{
\left(
-\frac{2q}{1-q}
+\frac{q^{\frac12}t^2}{1-q}
+\frac{q^{\frac12}t^{-2}}{1-q}
\right)
}_{\textrm{$1$ D3-brane}}. 
\end{align}

We note that due to the cancellations between bosonic and fermionic excitations 
the unflavored limit of the large $N$ half-index (\ref{largeN_u1_hindex}) and the single particle gravity index (\ref{ETW_1D3_gindex}) 
are insensitive to the fluctuation modes associated to an extra D3-brane 
as they become (\ref{largeN_Neu_u}) and (\ref{ETW_index1}) respectively. 

%U(N|2)
Similarly, for the interfaces between $U(N)$ and $U(2)$ gauge theories, 
the half-index in the unflavored limit can be expanded as
\begin{align}
\mathbb{II}_{\mathcal{N}}^{U(3)|U(2)}
&=1+3q^{\frac12}+7q+12q^{3/2}+17q^2+26q^{5/2}+35q^3+43q^{7/2}+\cdots,\\
\mathbb{II}_{\mathcal{N}}^{U(4)|U(2)}
&=1+3q^{\frac12}+7q+12q^{3/2}+20q^2+30q^{5/2}+41q^3+56q^{7/2}+\cdots,\\
\mathbb{II}_{\mathcal{N}}^{U(5)|U(2)}
&=1+3q^{\frac12}+7q+12q^{3/2}+20q^2+33q^{5/2}+45q^3+62q^{7/2}+\cdots,\\
\mathbb{II}_{\mathcal{N}}^{U(6)|U(2)}
&=1+3q^{\frac12}+7q+12q^{3/2}+20q^2+33q^{5/2}+48q^3+66q^{7/2}+\cdots. 
\end{align}
Again the finite $N$ correction appears at order $q^{\frac{N+1}{2}}$. 
In the large $N$ limit, the $U(N)|U(2)$ interface half-indices become 
\begin{align}
\mathbb{II}_{\mathcal{N}}^{U(\infty)|U(2)}&=
\mathbb{II}_{\mathcal{D}'}^{U(\infty)|U(2)}
=\prod_{n=1}^{\infty}
\frac{1-q^n}{1-q^{\frac{n}{2}}t^{-2n}}
\mathcal{I}^{U(2)}. 
\end{align}
Again the large $N$ interface half-index includes the large $N$ half-index (\ref{largeN_Neu}) of the half-BPS boundary condition 
describing the excitations on the $AdS_4$ ETW brane. 
Also it involves the Schur index $\mathcal{I}^{U(2)}$ of $\mathcal{N}=4$ $U(2)$ gauge theory. 

%U(N|M)
For general $U(N)|U(M)$ interface with $N>M$ 
when we take $N$ $\rightarrow$ $\infty$ 
while keeping $M$ finite, we get
\begin{align}
\label{large_uinf_um_hindex}
\mathbb{II}_{\mathcal{N}}^{U(\infty)|U(M)}&=
\mathbb{II}_{\mathcal{D}'}^{U(\infty)|U(M)}
=\prod_{n=1}^{\infty}
\frac{1-q^n}{1-q^{\frac{n}{2}}t^{-2n}}
\mathcal{I}^{U(M)}. 
\end{align}

When $N=M$, $N\rightarrow\infty$ and $M\rightarrow \infty$, the half-index is given by
%
\begin{comment}
The $U(N)|U(N)$ interface half-indices have the following expansions: 
\begin{align}
\mathbb{II}_{\mathcal{N}}^{U(1)|U(1)}
&=1+3q^{1/2}+2q+3q^{3/2}+5q^2-q^{5/2}+6q^3+6q^{7/2}+\cdots, \\
\mathbb{II}_{\mathcal{N}}^{U(2)|U(2)}
&=1+3q^{1/2}+7q+9q^{3/2}+13q^2+20q^{5/2}+22q^3+22q^{7/2}+\cdots, \\
\mathbb{II}_{\mathcal{N}}^{U(3)|U(3)}
&=1+3q^{1/2}+7q+16q^{3/2}+23q^2+36q^{5/2}+57q^3+75q^{7/2}+\cdots, \\
\mathbb{II}_{\mathcal{N}}^{U(4)|U(4)}
&=1+3q^{1/2}+7q+16q^{3/2}+32q^{2}+50q^{5/2}+81q^3+128q^{7/2}+\cdots, \\
\mathbb{II}_{\mathcal{N}}^{U(5)|U(5)}
&=1+3q^{1/2}+7q+16q^{3/2}+32q^{2}+61q^{5/2}+99q^3+160q^{7/2}+\cdots, \\
\mathbb{II}_{\mathcal{N}}^{U(6)|U(6)}
&=1+3q^{1/2}+7q+16q^{3/2}+32q^{2}+61q^{5/2}+112q^3+182q^{7/2}+\cdots.
\end{align}
\end{comment}
%
\begin{align}
\label{large_ns_hindex}
\mathbb{II}_{\mathcal{N}}^{U(\infty)|U(\infty)}
=\mathbb{II}_{\mathcal{D}'}^{U(\infty)|U(\infty)}
&=\prod_{n=1}^{\infty}
\frac{(1-q^n)^2}{(1-q^{\frac{n}{2}}t^{-2n})^2 (1-q^{\frac{n}{2}}t^{2n})}. 
\end{align}
Taking the plethystic logarithm, 
we obtain the corresponding single particle index that would capture the quantum fluctuations around the gravity dual geometry
\begin{align}
i^{\textrm{ETW$+AdS_5$}}
&=-\frac{2q}{1-q}+\frac{2q^{\frac12}t^{-2}}{1-q^{\frac12}t^{-2}}+\frac{q^{\frac12}t^{2}}{1-q^{\frac12}t^{2}}
\nonumber\\
&=\underbrace{\left(-\frac{q}{1-q}+\frac{q^{\frac12}t^{-2}}{1-q^{\frac12}t^{-2}}\right)}_{\textrm{ETW brane}}
+\underbrace{\left(-\frac{q}{1-q}+\frac{q^{\frac12}t^{2}}{1-q^{\frac12}t^{2}}+\frac{q^{\frac12}t^{-2}}{1-q^{\frac12}t^{-2}}\right)}_{AdS_5}. 
\end{align}
We see that the gravity index simply counts the quantum fluctuations of the ETW brane we obtained in (\ref{ETW_index1})
and those on $AdS_5\times S^5$ \cite{Kinney:2005ej}. 

When we turn off the fugacity $t$, we get 
\begin{align}
\mathbb{II}_{\mathcal{N}}^{U(\infty)|U(\infty)}
&=\mathbb{II}_{\mathcal{D}'}^{U(\infty)|U(\infty)}
=\frac{1}{(q)_{\infty}(q^{\frac12};q)_{\infty}^3}
\nonumber\\
&=\sum_{n\ge0}d_{\mathcal{N}}^{U(\infty)|U(\infty)}(n)q^{\frac{n}{2}}
\end{align}
According to the Meinardus theorem \cite{MR62781, MR1634067} and the convolution theorem \cite{MR3043606,MR3767503} 
\footnote{Also see \cite{Hayashi:2023txz} for the derivation of the convolution theorem for larger degeneracy 
including the subleading corrections based on the saddle point approximation. }, 
we find the asymptotic degeneracy of BPS local operators
\begin{align}
d_{\mathcal{N}}^{U(\infty)|U(\infty)}(n)
&\sim 
\frac{1}{4\sqrt{6}n}
\exp\left[
\frac{2\pi }{3^{\frac12}}n^{\frac12}
\right]. 
\end{align}

The normalized large $N$ half-index is 
\begin{align}
\mathcal{II}_{\mathcal{N}}^{U(\infty)|U(\infty)}
&=\prod_{n=1}^{\infty}(1+q^{\frac{n}{2}}), 
\end{align}
which agrees with (\ref{largeN_Neu_u}). 

In the Coulomb limit, the large $N$ half-index (\ref{large_ns_hindex}) becomes
\begin{align}
\label{large_unum_1/2}
{\mathbb{II}_{\mathcal{N}}^{U(\infty)|U(\infty)}}^{(C)}
&=\frac{1}{(\mathfrak{q}^2;\mathfrak{q}^2)_{\infty}^2}. 
\end{align}

%%%%%%%%%%%%%%%%%%%%%%%%%%%%%%%%%%%%%%%%%%%%%%%
\subsubsection{Giant graviton expansion}
%%%%%%%%%%%%%%%%%%%%%%%%%%%%%%%%%%%%%%%%%%%%%%%
Let us begin with the Coulomb limit of the interface half-indices. 
It follows from the closed-form expressions (\ref{unum_1/2}) and (\ref{large_unum_1/2}) that 
the $U(N)|U(M)$ interface half-index in the Coulomb limit has the following expansion: 
\begin{align}
\label{unum_halfggexp}
\frac{
{\mathbb{II}_{\mathcal{N}}^{U(N)|U(M)}}^{(C)}
}
{
{\mathbb{II}_{\mathcal{N}}^{U(\infty)|U(\infty)}}^{(C)}
}&=\sum_{m=0}^{\infty}\sum_{k=0}^m \mathfrak{q}^{2mM+2k(N-M)} 
\hat{F}_{m,k}^{(C)} (\mathfrak{q})
\end{align}
where 
\begin{align}
\label{unum_halfggindex}
\hat{F}_{m,k}^{(C)}(\mathfrak{q})
&=\frac{(-1)^m \mathfrak{q}^{m(m+1)-2k(m-k)}}{(\mathfrak{q}^2;\mathfrak{q}^2)_{k}(\mathfrak{q}^2;\mathfrak{q}^2)_{m-k}}. 
\end{align}
The expression (\ref{unum_halfggindex}) can be naturally identified with the giant graviton index 
that encodes the spectrum of the half-BPS giant gravitons in the $AdS_4$ bagpipe geometry 
with two asymptotic $AdS_5\times S^5$ regions and a single $5$-brane. 
The two indices $m$ and $k$ which take values in non-negative integers 
will correspond to the wrapping number of giants in the $AdS_5\times S^5$ regions with $M$ units of $5$-form flux, which we call $M$ D3 region, 
and that of giants of energy $(N-M)$ in the $AdS_4$ ETW brane region near the location of the $5$-brane respectively. 

When we put $m$ giants in the $M$ D3 region and $k$ giants in the ETW
region, there will be in principle, backreaction effects of the giants to the
flux. The probe giant adds one unit of flux, but does not change the
background metric. For the $k$ giants in the ETW region, there are $k$ units of flux coming from the ETW region 
and going out of the $M$ D3 region. In particular, there are $(m-k)$ units of excess flux, going through the five-cycle, transverse to the giants, and going out of the
$M$ D3 region so that $k$ ranges from $0$ to $m$. 
Due to the conservation of the flux flowing between the two different regions, 
$k$ units of flux, as the rest of the $m$ units of flux, balances the flux coming from the ETW region. 

In particular, when $k=0$ the giant graviton index (\ref{unum_halfggindex}) together with the factor $\mathfrak{q}^{2mM}$ 
takes the same form as (\ref{1/2gg_index}), 
which can be regarded as the index of the $m$ half-BPS giants in the $AdS_5\times S^5$ with $M$ units of $5$-form flux. 
In the expansion (\ref{unum_halfggexp}) 
the wrapping number $k$ of giants in the ETW brane near the location of the $5$-brane 
is bounded by the wrapping number $m$ of giants in the asymptotic $AdS_5\times S^5$ region. 
As discussed in section \ref{sec_gg_ETW}, 
the giant graviton index of the single half-BPS giant is obtained from the Landau problem. 
It can be obtained when $m=k=1$
\begin{align}
\hat{F}_{1,1}^{(C)}(\mathfrak{q})&=
-\frac{\mathfrak{q}^{2}}{1-\mathfrak{q}^2}. 
\end{align}

We observe that under $\mathfrak{q}\rightarrow \mathfrak{q}^{-1}$ 
the giant graviton index (\ref{unum_halfggindex}) becomes the interface half-index 
\begin{align}
\hat{F}_{m,k}^{(C)}(\mathfrak{q}^{-1})
&={\mathbb{II}_{\mathcal{N}}^{U(k)|U(m-k)}}^{(C)}(\mathfrak{q})
\nonumber\\
&=\langle W_{(1^k)}W_{\overline{(1^k)}}\rangle^{U(m)}_{\textrm{$\frac12$BPS}}(\mathfrak{q}^2), 
\end{align}
where $\langle W_{(1^k)}W_{\overline{(1^k)}}\rangle^{U(m)}_{\textrm{$\frac12$BPS}}(\mathfrak{q}^2)$ is the 
half-BPS limit of the two-point function of the antisymmetric Wilson line operators \cite{Hatsuda:2023imp}. 
Therefore the giant graviton expansion can be written as
\begin{align}
\label{unum_halfggexp2}
\frac{
{\mathbb{II}_{\mathcal{N}}^{U(N)|U(M)}}^{(C)}
}
{
{\mathbb{II}_{\mathcal{N}}^{U(\infty)|U(\infty)}}^{(C)}
}&=\sum_{m=0}^{\infty}\sum_{k=0}^m \mathfrak{q}^{2mM+2k(N-M)} 
\langle W_{(1^k)}W_{\overline{(1^k)}}\rangle^{U(m)}_{\textrm{$\frac12$BPS}}(\mathfrak{q}^{-2}). 
%{\mathbb{II}_{\mathcal{N}}^{U(m-k)|U(k)}}^{(C)}(\mathfrak{q}^{-1}). 
\end{align}

Note that the $U(N)|U(M)$ interface half-index in the Coulomb limit also has the following expansion: 
\begin{align}
\label{unum_halfggexp3}
\frac{
{\mathbb{II}_{\mathcal{N}}^{U(N)|U(M)}}^{(C)}
}
{
{\mathbb{II}_{\mathcal{N}}^{U(\infty)|U(\infty)}}^{(C)}
}&=\sum_{m_{1}=0}^{\infty}\sum_{m_2=0}^{\infty}
\mathfrak{q}^{2m_1 M+2m_2 N}
\langle W_{(1^{m_2})}W_{\overline{(1^{m_2})}}\rangle^{U(m_1+m_2)}_{\textrm{$\frac12$BPS}}(\mathfrak{q}^{-2}).
%{\mathbb{II}_{\mathcal{N}}^{U(m_1)|U(m_2)}}^{(C)}(\mathfrak{q}^{-1}). 
\end{align}
By contrast, the expansion (\ref{unum_halfggexp3}) contains a pair of non-negative integers $m_1$ and $m_2$ without any constraint. 
This may be naturally understood as alternative giant graviton expansions  
associated with the half-BPS giants with wrapping number $m_1$ and $m_2$ in the two asymptotic $AdS_5\times S^5$ regions. 

We are interested in studying the giant graviton expansions for the $U(N)|U(M)$ interface half-indices 
which can encode the spectra of the $1/8$-BPS giant gravitons in the $AdS_4$ bagpipe geometries. 

For $N=\infty$, we find from (\ref{large_uinf_um_hindex}) and (\ref{large_ns_hindex}) that 
\begin{align}
\frac{\mathbb{II}_{\mathcal{N}}^{U(\infty)|U(M)}}{\mathbb{II}_{\mathcal{N}}^{U(\infty)|U(\infty)}}(x;y)
&
=\sum_{m=0}^{\infty}x^{mM}\mathcal{I}^{U(m)}(x^{-1};xy), 
\label{eq:GGE-infty-M}
\end{align}
where we have used the variables $x$ and $y$ in (\ref{xy_var}). 
In this case, the giant graviton expansion is only associated with giants of energy $M$ in the asymptotic $AdS_5\times S^5$ region 
so that it is essentially same as the single sum giant graviton expansions for the $U(M)$ Schur index \cite{Gaiotto:2021xce}. 
Furthermore, it admits the following double sum giant graviton expansion: 
\begin{align}
\frac{\mathbb{II}_{\mathcal{N}}^{U(\infty)|U(M)}}{\mathbb{II}_{\mathcal{N}}^{U(\infty)|U(\infty)}}(x;y)
&
=\sum_{m_1=0}^{\infty}
\sum_{m_2=0}^{\infty}
x^{m_1M} y^{m_2M}(xy)^{m_1 m_2}\mathcal{I}^{U(m_1)}(x^{-1};xy)\mathcal{I}^{U(m_2)}(y^{-1};xy), 
\end{align}
which corresponds to the double sum giant graviton expansions for the $U(M)$ Schur index \cite{Arai:2020qaj}.

The general $U(N)|U(M)$ case is more involved, but we find an interesting structure. 
%We still expect that the half-index has a giant graviton expansion.
We have numerically checked that the half-index $\mathbb{II}_{\mathcal{N}}^{U(N)|U(M)}$ has the following expansion:
\begin{align}
\label{unum_gg1}
\frac{\mathbb{II}_{\mathcal{N}}^{U(N)|U(M)}}{\mathbb{II}_{\mathcal{N}}^{U(\infty)|U(\infty)}}(x;y)
&
=\sum_{m=0}^{\infty}\sum_{k=0}^{m}x^{mM+k(N-M)}
\hat{F}_{m-k,k}(x;y). 
\end{align}
Alternatively, we have 
\begin{align}
\label{unum_gg2}
\frac{\mathbb{II}_{\mathcal{N}}^{U(N)|U(M)}}{\mathbb{II}_{\mathcal{N}}^{U(\infty)|U(\infty)}}(x;y)
&
=\sum_{m_1=0}^\infty \sum_{m_2=0}^\infty x^{m_1 M+m_2 N} \hat{F}_{m_1, m_2}(x;y).
\end{align}
If we take the limit $N \to \infty$, the sum survives only for $m_2=0$.
Also it reduces to \eqref{eq:GGE-infty-M}. Therefore we conclude
\begin{align}
\hat{F}_{m, 0}(x;y)=\sigma_x[\mathcal{I}^{U(m)}(x;y)]. 
\end{align}
For the symmetry between $N$ and $M$ in (\ref{unum_gg2}), we also have 
\begin{align}
\hat{F}_{m_1, m_2}(x;y)=\hat{F}_{m_2, m_1}(x;y). 
\label{eq:symmetry-m1m2}
\end{align}
For small $(m_1, m_2)$, the numerical expansions of $\hat{F}_{m_1, m_2}(x;y)$ are given by
\begin{align}
\hat{F}_{1,1}(x;y)&=\frac{x^2}{(1-x)^2}+(1+2x)y+\biggl( \frac{1}{x^2}+\frac{2}{x}+1+3x^2 \biggr)y^2+\cdots,\\
\hat{F}_{2,1}(x;y)&=-\frac{x^4}{(1-x)^2(1-x^2)}-\frac{x(1+2x+2x^2)}{1-x^2}y\nonumber \\
&\hspace{2truecm}-\frac{1+2x+2x^2-x^4+3x^5+4x^6}{x^2(1-x^2)}y^2+\cdots\\
\hat{F}_{2,2}(x;y)&=\frac{x^6}{(1-x)^2(1-x^2)^2}+\frac{x^2(1+x+2x^2)}{(1-x)(1-x^2)}y\nonumber \\
&\hspace{2truecm}+\frac{1+x+2x^2-x^4-x^5+x^6+5x^7}{x^2(1-x)(1-x^2)}y^2+\cdots
\end{align}
We observe that these functions are exactly related to the unnormalized two-point functions of the antisymmetric Wilson lines, 
studied in \cite{Hatsuda:2023iwi, Hatsuda:2023imp}. 
We find that 
\begin{equation}
\begin{aligned}
\hat{F}_{m_1, m_2}(x;y)=\sigma_x [ \langle W_{(1^{m_2})}W_{\overline{(1^{m_2})}}\rangle^{U(m_1+m_2)}(x;y)],
\end{aligned}
\end{equation}
where for $m_2=0$, we have defined $\langle W_{(1^{0})}W_{\overline{(1^{0})}}\rangle^{U(m)}:=\mathcal{I}^{U(m)}$. 
The symmetry \eqref{eq:symmetry-m1m2} requires that the following equality for the correlators of the antisymmetric Wilson lines 
should hold: 
\begin{align}
\langle W_{(1^{m_2})}W_{\overline{(1^{m_2})}}\rangle^{U(m_1+m_2)}(x;y)
=\langle W_{(1^{m_1})}W_{\overline{(1^{m_1})}}\rangle^{U(m_1+m_2)}(x;y). 
\end{align}
In fact, this equality was observed in \cite{Hatsuda:2023iwi}.
Therefore, we conjecture that 
the $U(N)|U(M)$ interface half-index has the giant graviton expansion
\begin{align}
\frac{\mathbb{II}_{\mathcal{N}}^{U(N)|U(M)}}{\mathbb{II}_{\mathcal{N}}^{U(\infty)|U(\infty)}}(x;y)
&
=\sum_{m=0}^{\infty}\sum_{k=0}^{m}x^{mM+k(N-M)}
\sigma_x\left[
\langle W_{(1^{k})} W_{\overline{(1^k)}}\rangle^{U(m)}\rangle 
\right]. 
\end{align}

Furthermore, the important observation is that  
the double sum expansion of the two-point function of the antisymmetric Wilson lines is again given by 
the $U(N)|U(M)$ interface half-indices as the dual relation. 
Imamura and Inoue numerically computed the following brane sum expansion \cite{Imamura:2024pgp}: 
\begin{align}
\frac{\langle W_{(1^k)} W_{\overline{(1^k)}} \rangle^{U(N)}}
{\langle W_{(1^{\infty})} W_{\overline{(1^{\infty})}} \rangle^{U(\infty)}}
&=\sum_{m_1=0}^{\infty}\sum_{m_2=0}^{\infty}
x^{km_1+(N-k)m_2} \mathcal{F}_{m_1,0,m_2,0}(x;y)
\end{align}
with 
\begin{align}
 \mathcal{F}_{m_1,0,m_2,0}(x;y)&=
 \sigma_x 
 \mathbb{II}^{U(m_1|m_2)}_{\mathcal{N}}
 = \sigma_x 
 \mathbb{II}^{U(m_1|m_2)}_{\mathcal{D}'}.  
\end{align}
While this has been numerically checked in \cite{Imamura:2024pgp} 
by evaluating the NS5-type interface half-index (\ref{un_ns_um_hindex}) with $m_1+m_2\le 4$ up to the order $y^3$, 
we note that the S-dual D5-type interface half-index (\ref{un_d5_um_hindex}) and 
the formula (\ref{un_un_exp}) allow us to check higher powers of the expansions for larger values of $m_1$, $m_2$. 
\footnote{The condition $m_1+m_2\le 4$ for the NS5-type interface half-index is relaxed to $\min (m_1,m_2)\le 4$ for the D5-type interface half-index. }
Therefore, we conclude that 
the half-index of the $U(N)|U(M)$ interface in $\mathcal{N}=4$ SYM theory 
and the two-point function of Wilson lines in the rank-$k$ antisymmetric representation for $\mathcal{N}=4$ $U(m)$ SYM theory are dual with each other. 

%%%%%%%%%%%%%%%%%%%%%%%%%%%%%%%%%%%
\subsection*{Acknowledgements}
The work of Y.H. was supported in part by JSPS KAKENHI Grant Nos. 22K03641 and 23H01093. 
The work of H.L. was supported in part by National Key R\&D Program of China grant 2020YFA0713000, 
by Overseas high-level talents program, by Fundamental Research Funds for the Central Universities of China, and by Grant No. 3207012204. 
The work of T.O. was supported by the Startup Funding no.\ 4007012317 of the Southeast University. 
%%%%%%

\bibliographystyle{utphys}
\bibliography{ref}

\end{document}